\documentclass[aps,twocolumn]{revtex4}
\usepackage{color}
\usepackage{psfrag}
\usepackage{dcolumn}
\usepackage{bm}
\usepackage{float}
\usepackage[latin1]{inputenc}
\usepackage[spanish,english]{babel}
\usepackage{amsfonts}
\usepackage{amssymb,epsf}
\usepackage{graphicx}
\usepackage{slashed}
\usepackage{epstopdf}
\usepackage{amsmath,amssymb}
\usepackage{pdflscape}
\usepackage{adjustbox}
\usepackage{hyperref}
\usepackage{subcaption}
\begin{document}

\title{Influence of dark matter on the structure of strange quark stars in one-fluid model}

\author{J. Sedaghat\footnote{
		email address: J.sedaghat@shirazu.ac.ir}, G. H. Bordbar\footnote{
		email address: ghbordbar@shirazu.ac.ir (corresponding author)}, M. Haghighat\footnote{
		email address: m.haghighat@shirazu.ac.ir }, S. M. Zebarjad\footnote{
		email address: zebarjad@shirazu.ac.ir}}

\affiliation{Physics Department and Biruni Observatory, Shiraz University, Shiraz 71454, Iran}
	
\begin{abstract}	
This work studies the influence of scalar dark matter on the structural properties of strange quark stars (SQS) within a one-fluid framework, considering Yukawa interactions between dark matter and quark matter. Contributions from perturbative QCD, Yukawa interaction between scalar dark matter and quarks, and Bose-Einstein condensation of dark matter are included in the model. We first determine the allowable range of Yukawa interaction coupling by imposing the stability condition for strange quark matter (SQM). Using this range, we derive the equation of state (EOS) for different fractions of dark matter within the total pressure of SQS. These fractions are constrained by the tidal deformability limit from GW170817. The presence of dark matter alters the EOS, leading to changes in the mass-radius relationship, tidal deformability, and stability of SQS. We demonstrate that increasing the mass of dark matter softens the EOS, whereas higher fractions of dark matter lead to stiffer EOSs. We also explore the reasons behind this behavior. Our EOSs not only describe massive objects, such as PSR J0952-0607 and PSR J2215+5135, but also satisfy the tidal deformability constraint from GW170817. These results reveal that incorporating dark matter modifies the EOS, enabling the support of higher stellar masses while maintaining consistency with observational data.
 
\end{abstract}

\maketitle

\section{Introduction}
Dark matter is one of the universe's most mysterious elements, making up about $27\%$  of its total mass-energy content \cite{Martino2020,Planck Collaboration 2014,Persic1996,Bertone2018}. The standard model does not account for dark matter, even though its presence is inferred from gravitational effects on visible matter, such as the rotation curves of galaxies, gravitational lenses, and the large-scale structure of the cosmos \cite{Zwicky 1933,Allen2011,Frenk2012,Bennett2013}. We know that the nature of dark matter remains uncertain. One leading candidate is weakly interacting massive particles (WIMPs), which interact through gravity and possibly the weak nuclear force \cite{Weinberg1977,Goodman1985,Vysotskii1977,Adhikari2023}. Another possibility is axions, hypothetical particles that could solve the strong CP problem in quantum chromodynamics \cite{Weinberg1978,Preskill1983,Abbott1983}. Scalar dark matter, involving scalar fields, is also an intriguing option. Moreover, scalar fields could form a significant part of dark matter \cite{Matos2000,Matos2001,Bernal2006,Boehm2001,Gottel2024}. 

Studying compact stars with the consideration of dark matter introduces new opportunities to explore the characteristics of both dark matter and compact stars \cite{Karkevandi2022, Barbat2024, Lenzi2023, Pal2024, Anzuini2024}. The presence of dark matter within compact stars could modify their EOS, mass-radius relationship, dimensionless tidal deformability ($\Lambda$), and other characteristics \cite{Routaray2023,Das2019,Panotopoulos2017}. Furthermore, accounting for dark matter can affect the cooling rate of neutron stars \cite{Bhata2020}. Strange quark stars (SQS) \cite{Kurkela2010,Sedaghatplb2022a,Sedaghatplb2022b,Sedaghatepjc2024} are theoretical stars made up of strange quark matter (SQM), a type of matter believed to exist at very high densities \cite{Witten1984,Terazawa1989,Weber2005}. We investigate dark matter in relation to SQS by analyzing how a scalar field interacts with quark matter. This interaction can be represented using a scalar interaction in lagrangian, which introduces an extra term to the star's energy density and pressure. Scalar dark matter can impact the perturbative EOS of quark matter, resulting in different pressure-density relationships. This influence, may change the mass-radius relationship of SQS. Moreover, scalar dark matter could affect the stability of SQS, influencing whether they will eventually collapse into black holes or maintain their stability over time. Gravitational wave (GW) detectors, such as LIGO and Virgo have opened a new window into the study of compact stars and dark matter. The detection of GWs from binary mergers of neutron stars and black holes provides crucial information about the internal structure and composition of these objects \cite{Abbott2017,Abbott2018,Abbott2020,Abbott2020ApJL}. The gravitational wave signal from neutron star mergers carries information about the $\Lambda$, which is sensitive to the EoS. The presence of dark matter could alter the $\Lambda$, providing indirect evidence for its existence \cite{Das2019}. The observed merger rates and mass distributions of compact objects can constrain the properties of dark matter, as certain dark matter models predict specific merger outcomes \cite{Hamoud2017,Kavanagh2018}. The study of post-merger oscillations, known as ringdown, can reveal the internal structure of the remnant object. Dark matter could influence these oscillations, offering another potential signal \cite{Berti2009,Cardoso2019}. GW detectors could potentially discover exotic compact objects, such as dark matter stars or boson stars, which would provide direct evidence for the existence and properties of dark matter \cite{Cardoso2019,Liebling2022}. Incorporating dark matter into the study of compact stars depends on how dark matter interacts with ordinary (baryonic) matter. This can be divided into two scenarios: i) Dark matter particles that interact with baryonic matter (single fluid model): In this case, dark matter and baryonic matter interact, meaning their fluids are connected. They exchange energy and momentum, reaching a common stable state. As a result, a single TOV equation is used to describe the structure of the star \cite{Pal2024}. ii) Dark matter particles that do not interact with baryonic matter (two-fluid model): Here, dark matter and baryonic matter do not interact, so they do not exchange energy or momentum. Each type of matter follows its own EOS, representing its unique properties. Therefore, separate TOV-like equations are needed for each component, and these equations are connected only through the spacetime metric \cite{Pal2024,Panotopoulos2017}.

In this paper, we study the effects of dark matter on SQS by treating dark matter as a scalar field within SQM. We specifically focus on the role of the Yukawa interaction between dark matter and quark matter. When there is no interaction, dark matter's influence on the star's energy and pressure is accounted for by boson condensation, which requires a two-fluid model since dark matter and quark matter behave independently. However, when the Yukawa interaction is present, the quark matter and dark matter are no longer independent, leading to a single-fluid model. We investigate how this interaction affects the structural properties of SQS. 

The structure of paper is organized as follows. In section \ref{thep}, we calculate the thermodynamic potential ($\Omega$) of SQM in the presence of dark matter. This potential is divided into four components: the thermodynamic potential of free fermions, the potential from QCD interactions between quarks, the potential due to Yukawa interactions between quarks and scalar dark matter, and the potential from the condensation of dark matter particles.  In section \ref{EOSSA}, we determine the allowable values of the Yukawa interaction coupling constant, $\alpha_Y$, that satisfy the stability conditions for SQM. We then select a specific value of $\alpha_Y$ and analyze thermodynamic quantities, including the EOS, the speed of sound (ensuring causality), and the adiabatic index (indicating dynamical stability).  
Finally, in section \ref{SPOfSQS}, we investigate the structural properties of SQS in the presence of scalar dark matter for various dark matter masses and pressure fractions ($fr$) within SQM. Notably, we demonstrate that the QCD EOS alone cannot account for the massive objects PSR J0952-0607 and PSR J2215+5135. However, by incorporating the contributions of dark matter into the EOS, these objects can be explained as SQSs.

\section{calculation of thermodynamic potential}\label{thep}
The thermodynamic potential is an important concept in statistical mechanics and quantum field theory that describes the overall properties of a system in thermal equilibrium. It helps us to understand the relationships between key variables such as the temperature, pressure, energy density, and chemical potential \cite{kapusta}. In this paper, we calculate the thermodynamic potential using a field-theoretical method based on Feynman diagrams. The amplitude of these Feynman diagrams is directly related to the thermodynamic potential in statistical physics. By evaluating the diagrams, we can incorporate the effects of particle interactions and obtain a detailed description of the system's thermodynamic properties \cite{kapusta}. Here, the thermodynamic potential is divided into two parts, as shown in the following equation:
\begin{equation}\label{omegatotal}
\Omega = \Omega_{\text{free}}+\Omega_{\text{QCD}} + \Omega_{\text{Yukawa}}+\Omega_{\text{BEC}},
\end{equation}
where $\Omega_{\text{free}}$ and $\Omega_{\text{QCD}}$ represent the contributions from non-interacting fermions and QCD interaction between quarks respectively. Additionally, $\Omega_{\text{Yukawa}}$ accounts for the effects of the Yukawa interaction between quarks and scalar dark matter and $\Omega_{\text{BEC}}$ denotes the contribution from Bose-Einstein condensation of scalar dark particles. First, we discuss the parts $\Omega_{\text{free}}$ and $\Omega_{\text{QCD}}$, and then we will focus on $\Omega_{\text{Yukawa}}$ and $\Omega_{\text{BEC}}$ contributions.

\subsection{Perturbative QCD}

\begin{figure}
	\center{\includegraphics[width=2.5cm]
		{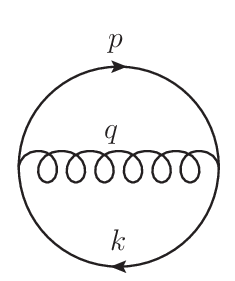}}
	\caption{\small{Two-loop diagram contributing to  perturbative part of  the grand potential at the leading order. }}
	\label{2loopQCD}
\end{figure}

$\Omega_{\text{free}}$ demonstrates the thermodynamic potential of free quarks and $\Omega_{\text{QCD}}$ represents the thermodynamic potential of the QCD interaction determined using a two-loop Feynman diagram (Fig. \ref{2loopQCD}). 
The calculation of these terms have been previously performed in \cite{Fraga2006} and \cite{Kurkela2010}, leading to the following expression:
\begin{equation}
	-\frac{\Omega_{free}+\Omega_{QCD}}{V}=\sum_{{N_f=1}}^{3}\left({\mathcal M}_1+\frac{{\mathcal M}_2\alpha_s(Q)}{4\pi}\right).
\end{equation}
Here, ${\mathcal M}_1$ and $\frac{{\mathcal M}_2 \alpha_s(Q)}{4\pi}$ represent the non-interacting and perturbative contributions, respectively. The expressions for ${\mathcal M}_1$ and ${\mathcal M}_2$ are given as follows:
\begin{equation}
	{\mathcal M}_1=
	\frac{N_c {\mu_f}^4}{24\pi^2}\bigg\{2\hat{u_f}^3-3z_f \hat{m_f}^2\bigg\} +\frac{{\mu_e}^4}{12\pi^2},\label{m1}
\end{equation}
\begin{equation}
	{\mathcal M}_2= \frac{d_A {\mu_f}^4}{4\pi^2}\Bigg\{-6z_f \hat{m_f}^2 \ln\frac{Q}{m} +2\hat{u_f}^4 - 4z_f \hat{m_f}^2 -3{z_f}^2 \Bigg\}, \label{m2}
\end{equation}
where $\hat{u_f}\equiv(\sqrt{{\mu_f}^2-m_f^2})/\mu_f$, $\hat{m_f}\equiv m_f/\mu_f$, $z_f \equiv \hat{u_f}-\hat{m_f}^2\,\ln\bigg[\frac{1+\hat{u_f}}{\hat{m_f}}\bigg]$ and $d_A\equiv {N_c}^2-1$ in which $N_c=3$ is the number of colors.
Additionally, $m_f$ and $\mu_f$ are the mass and chemical potential of a quark with flavor $f$, respectively and $\alpha_s$ is QCD running coupling constant given as follows \cite{Vermaseren}
\begin{equation} \label{QCD coupling}
\alpha_s (Q) =\frac{4 \pi  \left(1-\frac{2\beta_1 \log (L)}{{\beta_0}^2 L}\right)}{\beta_0L},
\end{equation}
where $\beta_0= 11 - 2\frac{N_f}{3}$, $\beta_1=51 - 19\frac{N_f}{3}$,  $L=2\log(\frac{Q}{\Lambda_{\overline{MS}}})$, and $N_f$ represents  the number of flavors, which we set to $3$. Here, $Q$ denotes the renormalization scale, and $\Lambda_{\overline{MS}}$ represents the renormalization point in the minimal subtraction scheme.  Following the results for stable SQM from \cite{Kurkela2010}, we set $Q = \frac{4}{3}(\mu_u + \mu_d + \mu_s)$. $\Lambda_{\overline{MS}}$ is extracted from the particle data group 2023 dataset \cite{Workman} by matching $\alpha_s(m_{\tau})$ to the value $0.314^{+0.014}_{-0.014}$. The tau particle mass, $m_{\tau}$, is taken as $1776.86 \, \text{MeV}$ \cite{Workman} (for more discussion see \cite{sedaghatannals}). In the present work, the masses of up and down quarks are assumed to be zero, while the running mass of the strange quark, denoted as $m_s(Q)$, is given by: 
\cite{Vermaseren,Fraga2006}.
\begin{equation}
m_{s}(Q)=m_{s}(2GeV)\left[ \dfrac{\alpha _{s}(Q)}{\alpha _{s}(2GeV)}\right]
^{\dfrac{\gamma _{0}}{\beta _{0}}},  \label{9}
\end{equation}%
where, $\gamma_0$, the anomalous dimension, is expressed as $3\dfrac{N_c^2 - 1}{2N_c}$. Based on the latest particle data group results, the mass of the strange quark at $2\,\text{GeV}$, denoted by $m_s(2\,\text{GeV})$, is reported as $93.4^{+8.6}_{-3.4}\,\text{MeV}$. For our calculations, we use the central value of this range, $m_s(2\,\text{GeV}) = 96\,\text{MeV}$ (see \cite{sedaghatannals} for additional details).

\subsection{Yukawa interaction}

\begin{figure}
	\center{\includegraphics[width=2.5cm]
		{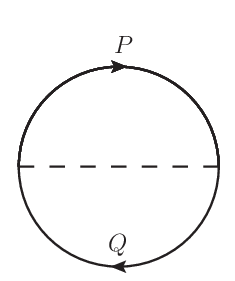}}
	\caption{\small{Two-loop diagram related to Yukawa interaction between quarks and dark matter. }}
	\label{2loopYukawa}
\end{figure}

{Yukawa-type interactions between a neutral scalar field and quarks naturally emerge in well-motivated extensions of the Higgs sector beyond the Standard Model. These extensions introduce additional scalar degrees of freedom, which can serve as dark matter candidates. Among the various Higgs sector, appropriate modifications that allow such interactions, the two-Higgs-doublet model (2HDM) and Higgs triplet models are two prominent examples. In these frameworks, extra scalar fields arise, enabling Yukawa couplings to quarks. In these models, after spontaneous symmetry breaking, the physical Higgs eigenstates mix with the new scalar fields, leading to effective interactions of the form $-g\phi\overline{\psi}\psi$, where $\phi$ is a neutral scalar and $\phi$ is the fermion field. If this scalar is also a dark matter candidate, its interaction with quarks can be understood as a natural consequence of Higgs mixing effects. These extensions are well explored in the literature ( for example, see \cite{Branco2012} for 2HDM and \cite{Gunion1989} for Higgs triplet models). Our aim in this paper is to investigate how this type of Yukawa interaction influences the structural properties of SQS.}
\\
The term $\Omega_{\text{Yukawa}}$ in Eq. (\ref{omegatotal}) represents the thermodynamic potential arising from the Yukawa interaction between scalar dark matter and quarks. To compute $\Omega_{\text{Yukawa}}$, we apply a powerful approach outlined in Ref. \cite{Ghisoiu2017}, commonly referred to as the cutting rules. This technique transforms the evaluation of Feynman integrals at finite chemical potential and zero temperature into the calculation of three-dimensional phase space integrals for on-shell amplitudes at $T = \mu = 0$. Using this method, the $N$-loop, $n$-point, one-particle irreducible Feynman diagram, $F(P_k, \mu)$, can be rewritten as:
\begin{align}
	F(P_k,\mu)=&F_{0- cut}(P_k)\\ \notag+ &F_{1-cut}(P_k,\mu)+...+F_{N-cut}(P_k,\mu).
\end{align}
In this approach, $F_{0-\text{cut}}(P_k)$ refers to the initial diagram calculated at zero chemical potential, while the additional terms arise from the cutting procedure. Specifically, $F_{j-\text{cut}}(P_k, \mu)$ represents the sum of all diagrams where $j$ internal fermion propagators have been cut from the original diagram. Below, we describe the steps involved in the cutting procedure:
\begin{enumerate}
\item The cut propagators are eliminated from the initial graph.
\item The amplitude for the remaining $N-j$ loops and $n+2j$ points is determined under the condition that all external momenta are real. 
\item The cut momenta $p_i$ are put  on-shell by assigning $({p_0})_i = iE_i$.
\item The final expression is evaluated by integrating over the three-dimensional cut momenta $p_i$, weighted by $\dfrac{-\theta(\mu-E_{i})}{2E_{i}}$. Here,  for a free quark of flavor $f$ and mass $m_f$, the energy $E_p$ is expressed as $E_p = \sqrt{p^{2} + m_{f}^{2}}$.
\end{enumerate}
The Yukawa interaction Lagrangian between the quark fields ($\psi$) and the scalar dark matter field ($\phi$) is expressed as  $-g\phi\overline{\psi}\psi$. The thermodynamic potential related to this interaction is calculated using a two-loop diagram, as shown in Fig. \ref{2loopYukawa}. In this diagram, the thick lines represent quark propagators, while the dashed line corresponds to the dark scalar propagator. In Euclidean space, these propagators take the forms  $D(P)=\frac{-i\slashed{p}}{P^2+m^2}$ for quarks and $\frac{1}{K^2+m^2}$ for the dark scalar field. Notably, in Euclidean space at finite chemical potential, the squared momentum is given by $P^2 = (P_0 + i\mu)^2 + p^2$ \cite{Kurkela2010}. The Feynman rule for each vertex in this interaction is expressed as $g = \sqrt{4\pi\alpha_Y}$.
Incorporating the Feynman rules in euclidean space, we get,
\begin{align}\label{omegaYukawa2}
	\dfrac{\Omega _{\text{Yukawa}}}{V}&=4\pi\alpha_Y \int \dfrac{d^{4}P}{(2\pi )^{4}}\int \dfrac{
		d^{4}Q}{(2\pi )^{4}}\dfrac{1}{(P-Q)^{2}}  \notag 
	\notag \\
	&\times  Tr\left[ D(P)D(Q)\right] \notag \\
	&=4\pi\alpha_Y\int \dfrac{d^{4}P}{(2\pi )^{4}}\int \dfrac{
		d^{4}Q}{(2\pi )^{4}}\dfrac{1}{(P-Q)^{2}}  \notag 
	\notag \\
	&\times \frac{Tr\left[(-i\slashed{P}+m_f)(-i\slashed{Q}+m_f)   \right]}{[(P_0+i\mu)^2+E^2_p][(Q_0+i\mu)^2+E^2_q][P-Q]^2}.
\end{align}
Using the cutting rules leads to
\begin{align}
	\dfrac{\Omega _{\text{Yukawa}}}{V}=F_{0-cut}+F_{1-cut}+F_{2-cut},
\end{align}
Since $ F_{0-cut} $ is independent of the chemical potential, it does not contribute to the quark number densities and it is therefore omitted. Finally, multiplying Eq. (\ref{omegaYukawa2}) by a factor of \(1/2\), which arises from the perturbative expansion \cite{kapusta}, we obtain:
\begin{align}
	F_{1-cut}&=-4\pi\alpha_Y\int \frac{\Theta(\mu-E_p)}{2E_p} \dfrac{d^{3}p}{(2\pi )^{3}}\int \dfrac{
		d^{4}Q}{(2\pi )^{4}}(m^2_f-P.Q)  \notag 
	\notag \\
	&\times \frac{4}{[Q^2_0+E^2_q][(P_0-Q_0)^2+(p-q)^2+m^2_{\text{DM}}]} \arrowvert_{P_0=iE_p}
\end{align}
\begin{align}
	F_{2-cut}&=2\pi\alpha_Y\int \frac{\Theta(\mu-E_p)}{2E_p} \dfrac{d^{3}p}{(2\pi )^{3}}\int \frac{\Theta(\mu-E_q)}{2E_q}\dfrac{
		d^{3}q}{(2\pi)^{3}}  \notag 
	\notag \\
	&\times \frac{4(m^2_f-P.Q)}{(P_0-Q_0)^2+(p-q)^2+m^2_{\text{DM}}} \arrowvert\substack{P_0 = iE_p \\ Q_0 = iE_q}
\end{align}
To regularize the integral without $\Theta$ function, we apply a cut-off to the upper limit of the integral. This cut-off is defined as the maximum value of renormalization scale, approximately on the order of 2 GeV \cite{Kurkela2010}. 
Based on the constraints on the dark matter self-interaction cross-section derived in Refs. \cite{Panotopoulos2017b,Lopes2018}, we consider three values for $m_{\text{DM}}$: $50 \, \text{MeV}$, $100 \, \text{MeV}$, and $160 \, \text{MeV}$. We then investigate the structural properties of SQSs in the presence of dark matter. 

\subsection{Bose-Einstein condensation of dark matter}

When the thermal wavelength becomes larger than the average distance between particles, particles in a dilute Bose gas merge into a single quantum state, which is known as a Bose-Einstein condensation (BEC). Therefore, under the assumption of absolute zero temperature ($T = 0  K$), the majority of dark matter particles settle into the condensed phase. In this state, interactions are dominated by low-energy collisions, which can be described using the s-wave scattering length $l_a$, without needing to consider the specifics of the two-particle forces. Hence, the complex interaction potential is simplified to an equivalent repulsive force as follows \cite{Panotopoulos2017b,Lopes2018,LopesDas}:
\begin{equation}
V(\textbf{r}
-\textbf{r}')=\frac{4\pi l_a}{m_{\text{DM}}}\delta(\textbf{r}
-\textbf{r}')
\end{equation}
The EOS related to this condensation has been derived in Refs. \cite{Panotopoulos2017b,Lopes2018} which is given by:
\begin{equation}\label{omegaBEC}
-\frac{\Omega_{BEC}}{V}=P_{BEC}=\frac{2\pi l_a}{m_{\text{DM}}^3}\epsilon_{BEC}^2
\end{equation}
Here, $l_a$, the s-wave scattering length, is taken as 1 $fm$ \cite{Panotopoulos2017b,Lopes2018,LopesDas}. In our calculations, the contributions of $P_{BEC}$ and $\epsilon_{BEC}$ are incorporated into the pressures and energy densities associated with non-interacting quarks, QCD, and Yukawa interactions. We examine various $fr$s of dark matter in the total pressure of the star to determine the structural properties of  SQS. The maximum $fr$ is selected to comply with the constraint $\Lambda_{1.4M_\odot} < 580$. It must be noted that to determine the thermodynamic potential, the value of $ \alpha_Y $ must be specified. In the next section, we will determine the allowed values of $ \alpha_Y $ using the stability condition of SQM. Then, having the thermodynamic potential, we will obtain the thermodynamic properties of SQM. These properties include the EOS, sound speed, and adiabatic index. 
  
\section{Equation of state, speed of sound and adiabatic index}\label{EOSSA}

In this section, we use the thermodynamic potential derived in the previous section to determine the thermodynamic properties of SQM. We first derive the EOSs based on specific physical conditions, which will be described later. Subsequently, we calculate the sound speed and adiabatic index to demonstrate that the EOSs satisfy the causality condition and ensures dynamical stability.

\subsection{Equation of state}

 The pressure, quark number density, and energy density can be derived from the thermodynamic potential using the following relations,
\begin{equation}
P=-B_{eff}-\frac{\Omega }{V},
\end{equation}
\begin{equation}
n_f=\frac{\partial P}{\partial \mu_f},
\end{equation}
and
\begin{equation}\label{EOS}
\epsilon=\sum_{{N_f}}\mu_fn_f+\mu_en_e-P,
\end{equation}
where the quark number density for flavor $f$ is denoted by $n_f$, and the electron number density, $n_e$, is determined by the formula $n_e = \frac{\mu_e^3}{3\pi^2}$. The effective bag constant, $ B_{\text{eff}} $, represents non-perturbative contributions not captured by the weak coupling expansion. Notably, the bag constant in the MIT bag model accounts for all interactions between quarks, whereas $ B_{\text{eff}} $ in the QCD perturbative expansion is included alongside the perturbative terms. This distinction results in $ B_{\text{eff}} $ having a significantly lower value than the MIT bag model \cite{sedaghatannals}. In the perturbative QCD EOS, $ B_{\text{eff}} $ is chosen to ensure that the pressure vanishes at the surface of the star \cite{Kurkela2010,Sedaghatplb2022a,Sedaghatplb2022b,sedaghatannals}. In our calculations, we consider $ B_{\text{eff}} $ to be $ 50 \, \frac{\text{MeV}}{\text{fm}^3} $. To obtain $\Omega_{\text{Yukawa}}$, we must determine the value of $\alpha_{Y}$. For each value of $ m_{\text{DM}} $, we identify the range of $ \alpha_Y $ values that satisfy the stability condition for SQM. If SQM exists, it serves as the true ground state of QCD, implying that the energy per baryon in SQM at vanishing pressure is lower than that of the most stable nuclear matter. Therefore, considering the baryon number density $ n_B = \frac{n_u + n_d + n_s}{3} $, the condition $ \frac{\epsilon}{n_B} < 930 \, \text{MeV} $ must be enforced \cite{Weber2005}.
\begin{figure}[h!]
	\centering
	\includegraphics[width=8cm]{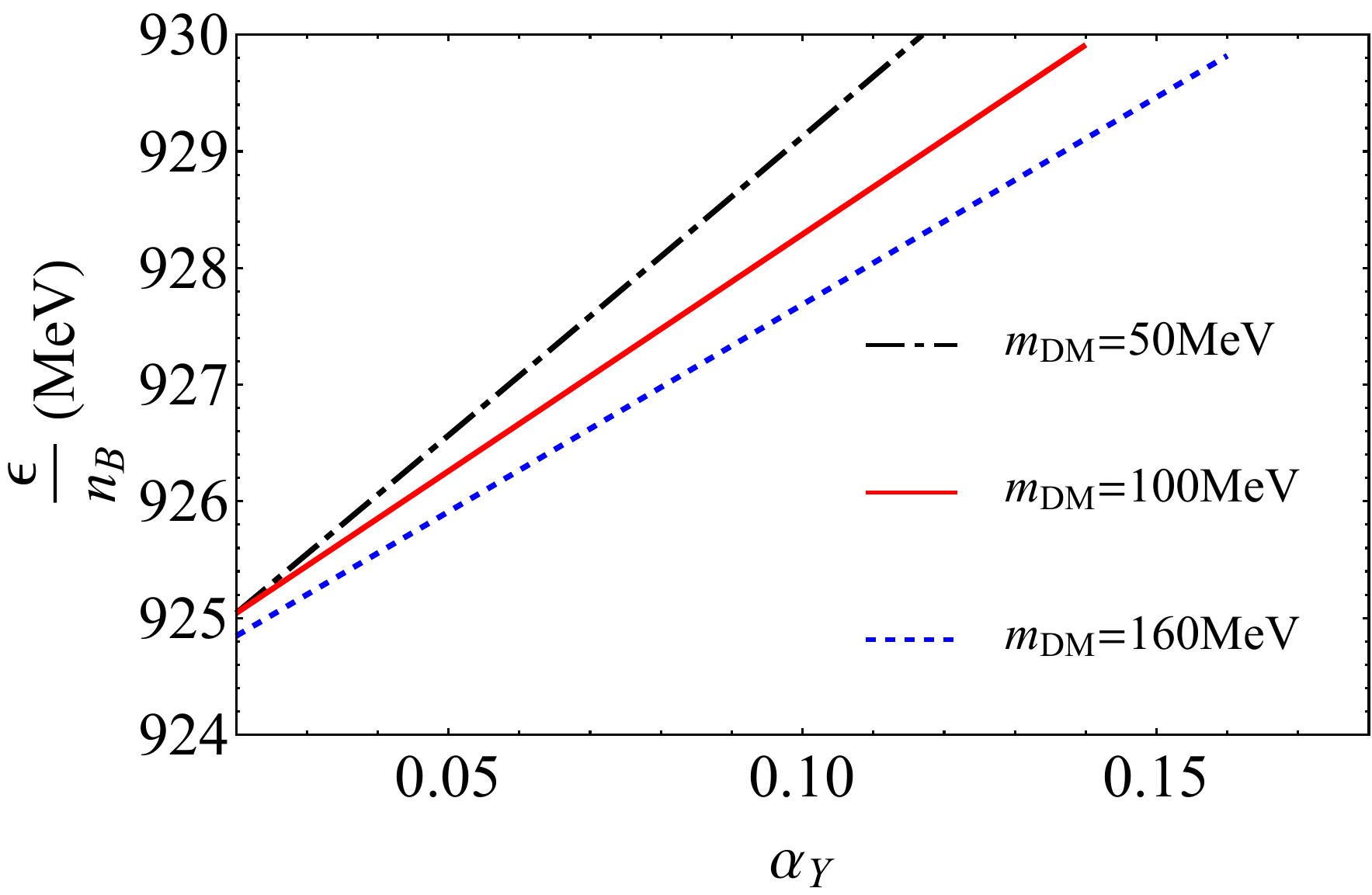}
	\caption{Energy per baryon for different values of $\alpha_Y$ and $m_{\text{DM}}$.}	\label{stabilitycondition}
\end{figure}
Fig. \ref{stabilitycondition} illustrates the variation of baryon energy density with respect to $ \alpha_Y $. It is evident that as $ \alpha_Y $ increases, $ \frac{\epsilon}{n_B} $ also rises.  Table \ref{alphamax} presents the maximum permissible values of $ \alpha_Y $ for each corresponding $ m_{\text{DM}} $. As indicated, for $ m_{\text{DM}} = 50 \, \text{MeV} $, the maximum allowable $ \alpha_Y $ is approximately $0.11$. This value increases to about $0.14$ for $ m_{\text{DM}} = 100 \, \text{MeV} $, and further rises to around $0.16$ for $ m_{\text{DM}} = 160 \, \text{MeV} $.
\begin{table}[h!]
	\caption{{\protect\small {maximum permissible values of $ \alpha_Y $ for each corresponding $ m_{\text{DM}} $}}}
	\label{alphamax}\centering
	\par
	\begin{adjustbox}{width=.2\textwidth}
		\begin{tabular}{|c|c|c|c|c|c|c|}
\hline
\scriptsize$m_{\text{DM}}(\text{MeV})$ & \scriptsize ${\alpha_{Y}}_{max}$\\	\hline
			\scriptsize$50$ &\scriptsize 0.11  \\ \hline
		\scriptsize	$100$ & \scriptsize0.14  \\ \hline
		\scriptsize	$160$ & \scriptsize 0.16 \\ \hline
		\end{tabular}
	\end{adjustbox}
\end{table}
Hereafter, among the permissible values of $ \alpha_Y $, we select $ \alpha_Y = 0.1 $ to calculate the thermodynamic properties of SQM in the presence of dark matter. Given the thermodynamic potential $\Omega$, the relationship between pressure and energy density can be determined using Eq. (\ref{EOS}), which is known as EOS. The allowed values of the chemical potentials for the quarks are determined by applying the following three physical conditions,
\\
i) Charge neutrality that requires that the system remains electrically neutral on a locally, which implies:
\\
\begin{equation}
\frac{2}{3}n_{u}-\frac{1}{3}n_{d}-\frac{1}{3}n_{s}-n_{e}=0,  \label{charge neutrality}
\end{equation}
ii) Beta equilibrium that occurs when the chemical potentials are balanced through weak interactions \cite{Kurkela2010}:
\begin{equation*}
d\longrightarrow u+e+\bar{\nu}_{e},~~~\&~~~u+e\longrightarrow d+\nu _{e},
\end{equation*}%
\begin{equation*}
s\longrightarrow u+e+\bar{\nu}_{e},~~~\&~~~u+e\longrightarrow s+\nu _{e},
\end{equation*}%
\begin{equation}
s+u\longleftrightarrow d+u,
\end{equation}%
which result in the following constraints for the chemical potentials of the quarks.
\begin{equation}
\mu _{s}=\mu _{d}\equiv \mu ,~~~\&~~~\mu _{u}=\mu -\mu _{e}.  \label{24}
\end{equation}
iii) The baryon number density, $ n_B = \frac{n_u + n_d + n_s}{3} $, must be greater than the saturation density to ensure the quark matter is physically realistic. Once the chemical potentials that satisfy these three conditions are determined, the following physical conditions on pressure and energy density must also be imposed.
\begin{enumerate}
\item The pressure should remain positive throughout the system and decrease to zero at the boundary of the star.
\item To ensure stability, the minimum energy per baryon must be below $930 \text{MeV}$, which is the binding energy of the most stable nuclei.
\item To respect causality, the speed of sound, given by $\sqrt{\frac{dp}{d\epsilon}}$, must not exceed the speed of light.
\item The adiabatic index, $\Gamma =\frac{dP}{d\epsilon} \dfrac{(P+\epsilon )}{P}$, must exceed $ \frac{4}{3} $ to ensure dynamical stability.
\item The pressure and energy density must meet to the conditions $ \epsilon > 0 $ and $ \epsilon > P $ for physical consistency.
\end{enumerate}
\begin{figure}[h!]
	\centering
	\begin{subfigure}{0.45\textwidth}
		\centering
		\includegraphics[width=\textwidth]{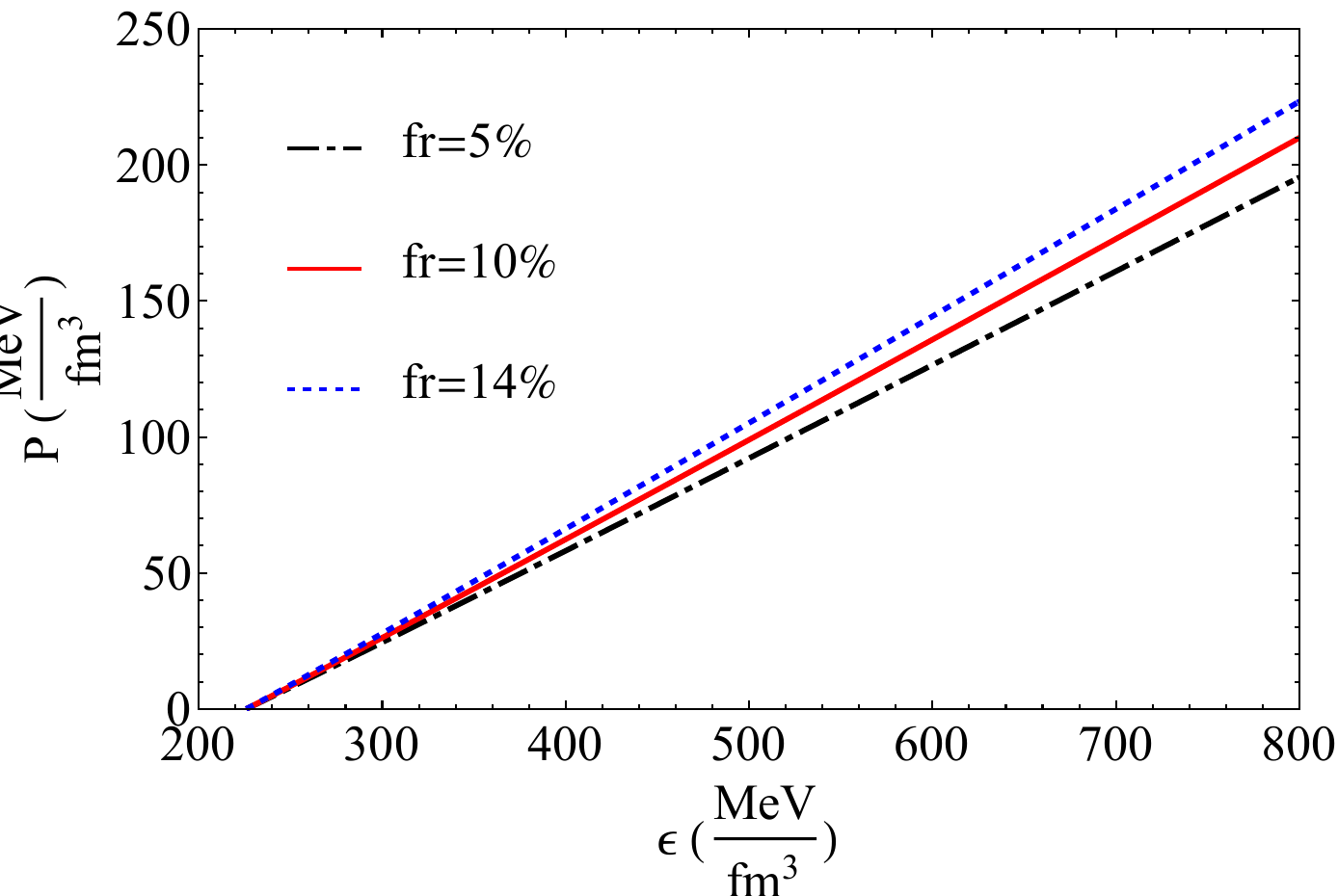}
		{$m_{\text{DM}}$=50 MeV}
	\end{subfigure}
	\hfill
	\begin{subfigure}{0.45\textwidth}
		\centering
		\includegraphics[width=\textwidth]{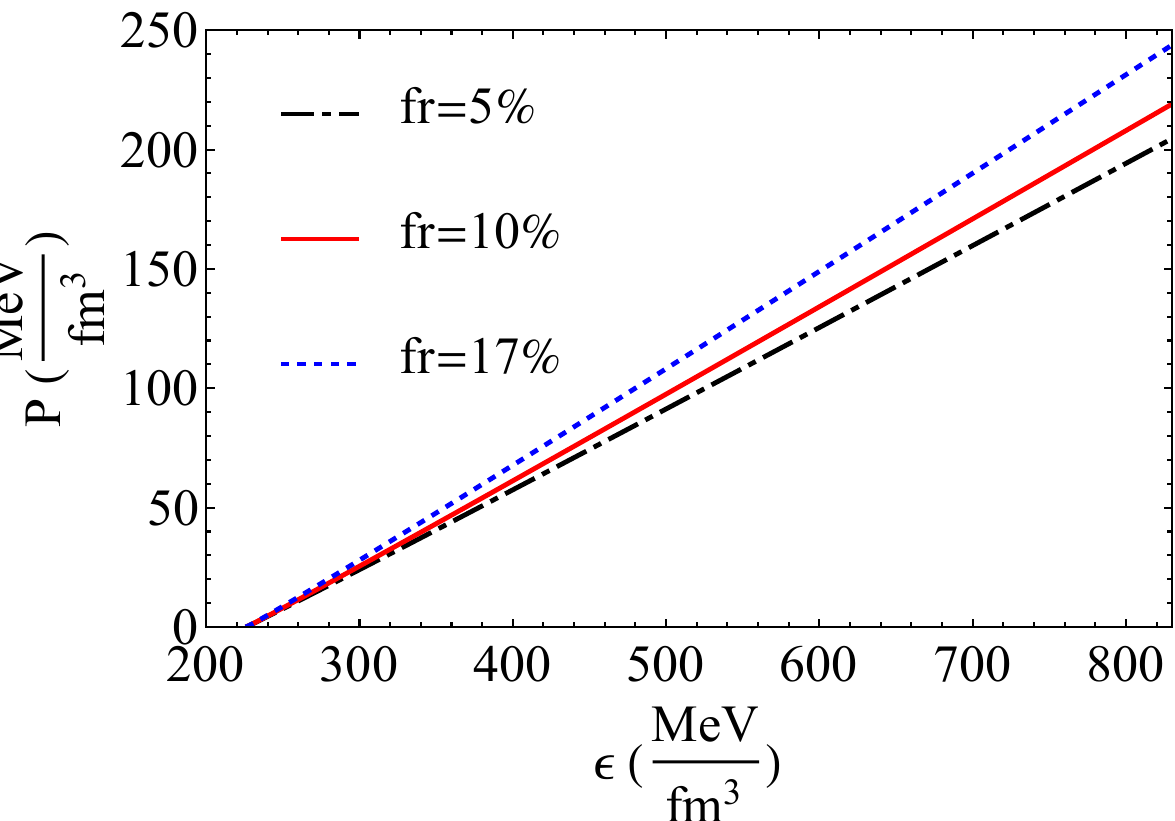}
		{$m_{\text{DM}}$=100 MeV}
	\end{subfigure}
	\hfill
	\begin{subfigure}{0.45\textwidth}
		\centering
		\includegraphics[width=\textwidth]{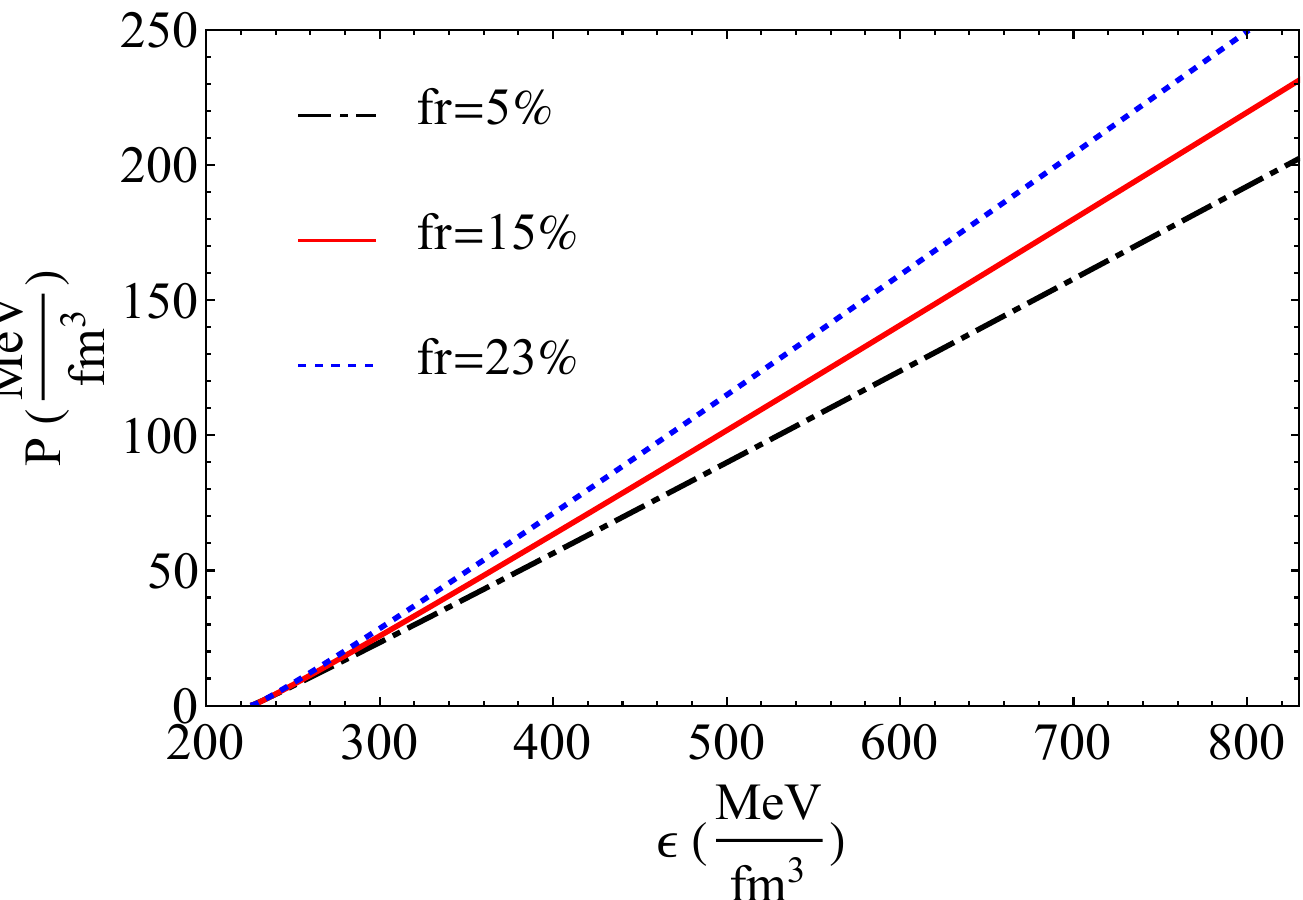}
		{$m_{\text{DM}}$=160 MeV}
	\end{subfigure}
	\caption{EOS for different values of $m_{\text{DM}}$ and $fr$.}
	\label{different EOSs}
\end{figure}

Fig. \ref{different EOSs} shows the EOS of SQM for different values of $m_{\text{DM}}$ and $fr$ of dark matter. For each value of $m_{\text{DM}}$, as the $fr$ increases, the EOS becomes stiffer, meaning the pressure increases more steeply with energy density which can support higher stellar masses. It is well known that stiffer EOS leads to the higher $\Lambda$ \cite{Sedaghatepjc2024}. We adjust $fr$ to ensure compliance with the $\Lambda$ constraint from GW170817. Further details will be provided in section \ref{SPOfSQS} . As observed, for a fixed $fr$ ($fr = 5\%$), the EOS becomes stiffer for smaller values of $m_{\text{DM}}$. This shows that by increasing the dark matter mass, EOS becomes softer. The increase in maximum allowable $fr$ of dark matter by increasing $m_{\text{DM}}$ (14\% for $ m_{\text{DM}} = 50 $ MeV, 17\% for $ m_{\text{DM}} = 100 $ MeV, and 23\% for $ m_{\text{DM}} = 160 $ MeV) illustrate this behavior. 
Here, it should be noted that the Yukawa interaction softens the EOS, while the BEC of dark matter stiffens it. Consequently, a higher BEC contribution results in a stiffer EOS, whereas a lower BEC contribution leads to a softer EOS. Increasing $ fr $ at a constant $ m_{\text{DM}} $ enhances the BEC contribution, making the EOS stiffer. Conversely, increasing $ m_{\text{DM}} $ at a fixed $ fr $ reduces the BEC contribution (see Eq. (\ref{omegaBEC})), causing the EOS to soften.

\subsection{Sound speed}

\begin{figure}[h!]
	\centering
	\begin{subfigure}{0.463\textwidth}
		\centering
		\includegraphics[width=\textwidth]{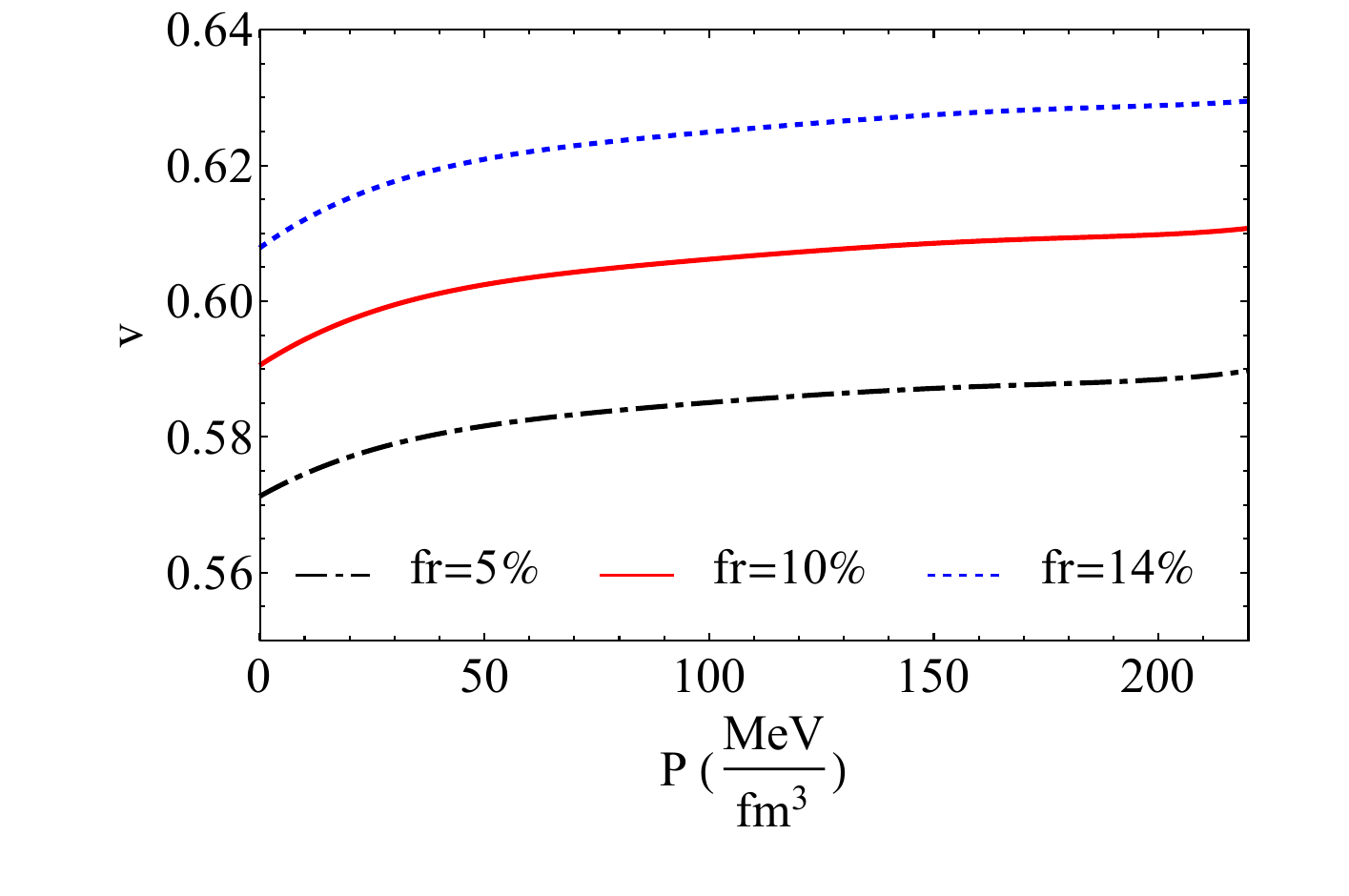}
		{$m_{\text{DM}}$=50 MeV}
	\end{subfigure}
	\hfill
	\begin{subfigure}{0.45\textwidth}
		\centering
		\includegraphics[width=\textwidth]{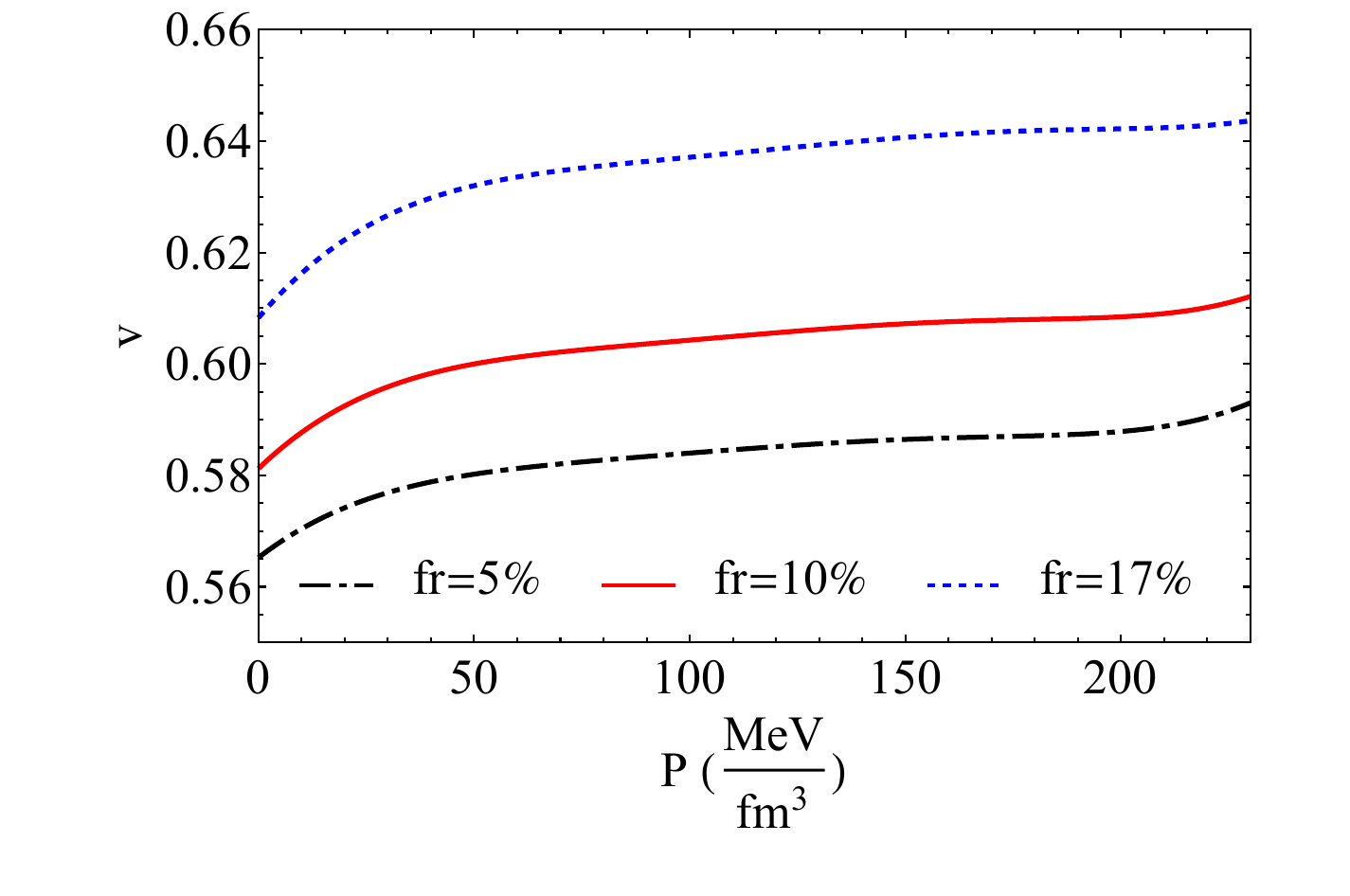}
		{$m_{\text{DM}}$=100 MeV}
	\end{subfigure}
	\hfill
	\begin{subfigure}{0.45\textwidth}
		\centering
		\includegraphics[width=\textwidth]{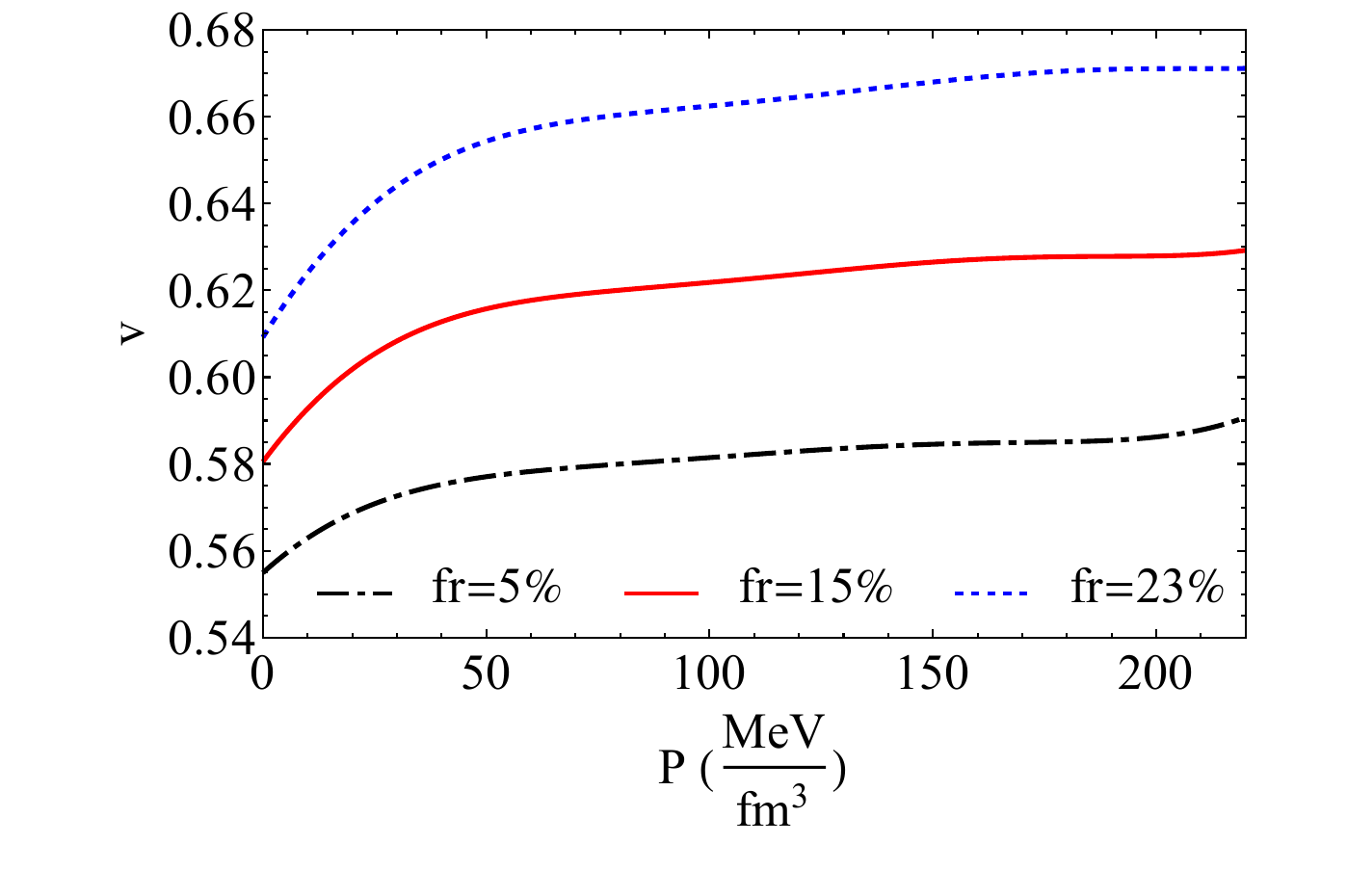}
		{$m_{\text{DM}}$=160 MeV}
	\end{subfigure}
	\caption{Speed of sound (in the unit of light speed $c$) for different values of $m_{\text{DM}}$ and $fr$.}
	\label{different sound speeds}
\end{figure}

For an EOS to be appropriate for compact stars, it must meet the fundamental requirement that no signal within the star can propagate faster than the speed of light. This is expressed through the condition $\frac{v}{c}\leq1$, where $v$ represents the speed of sound ($v=\sqrt{\frac{dP}{d\epsilon}}$). The violation of this condition would conflict with the principles of special relativity, leading the model unphysical. Fig. \ref{different sound speeds} illustrates the speed of sound as a function of pressure. It is evident that all diagrams satisfy this requirement, confirming the physical viability of the EOSs. Furthermore, the results confirm that increasing the $fr$ stiffens the EOS, leading to higher sound speeds while maintaining causality. Moreover, the results show that larger dark matter mass softens the EOS at fixed $fr$, resulting in slightly lower sound speeds.

\subsection{Adiabatic index}

\begin{figure}[h!]
	\centering
	\begin{subfigure}{0.45\textwidth}
		\centering
		\includegraphics[width=\textwidth]{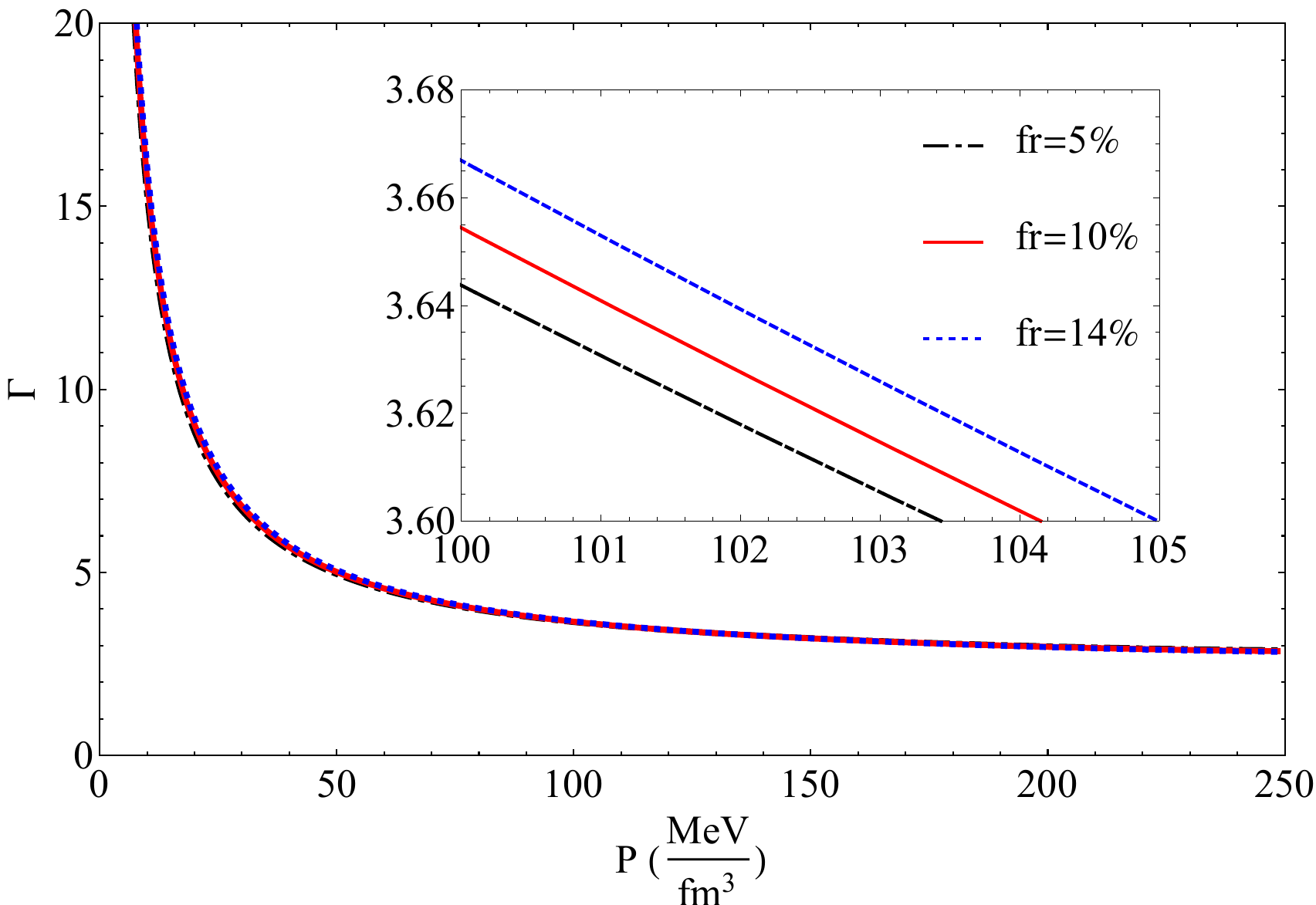}
		{$m_{\text{DM}}$=50 MeV}
	\end{subfigure}
	\hfill
	\begin{subfigure}{0.45\textwidth}
		\centering
		\includegraphics[width=\textwidth]{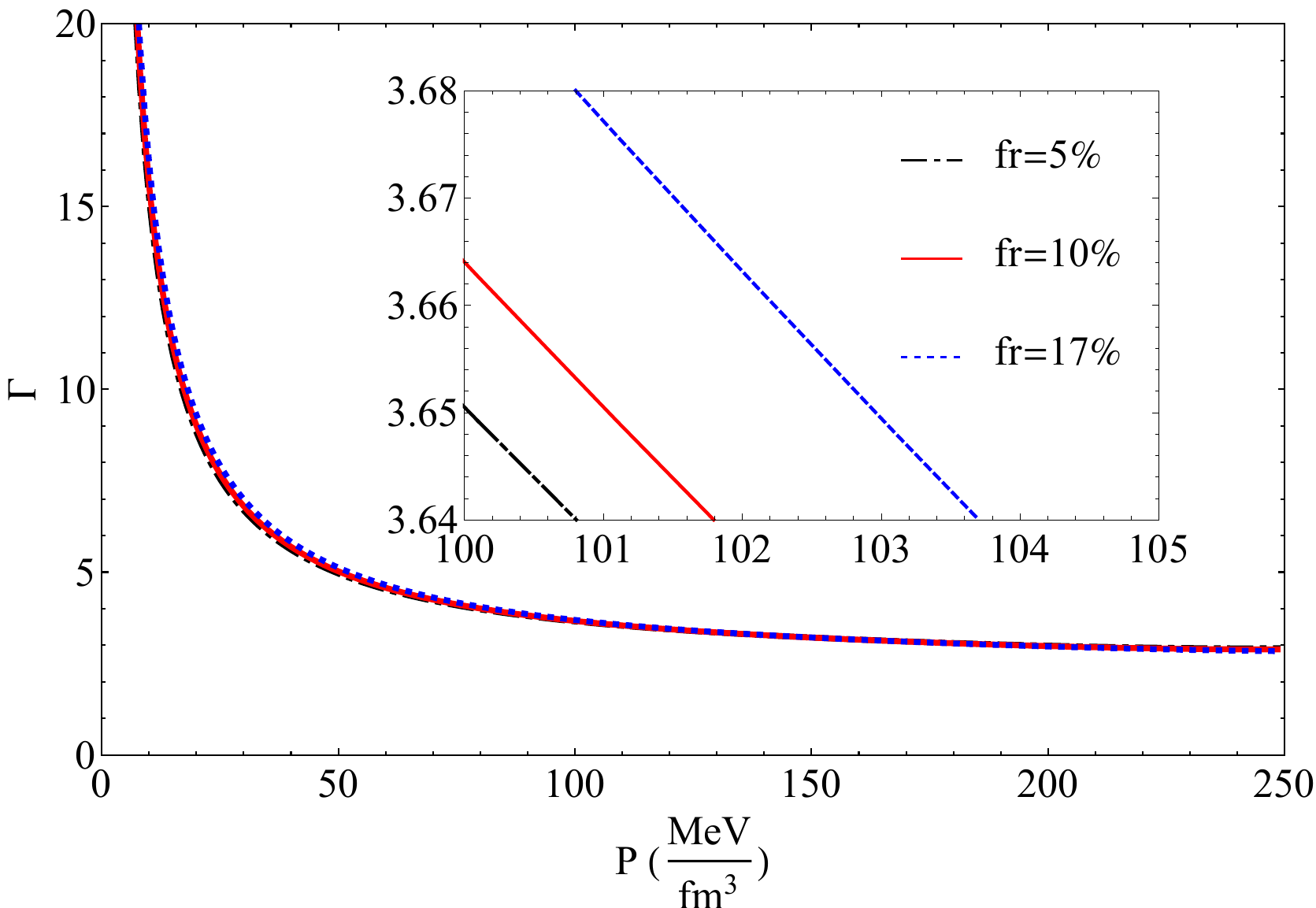}
		{$m_{\text{DM}}$=100 MeV}
	\end{subfigure}
	\hfill
	\begin{subfigure}{0.45\textwidth}
		\centering
		\includegraphics[width=\textwidth]{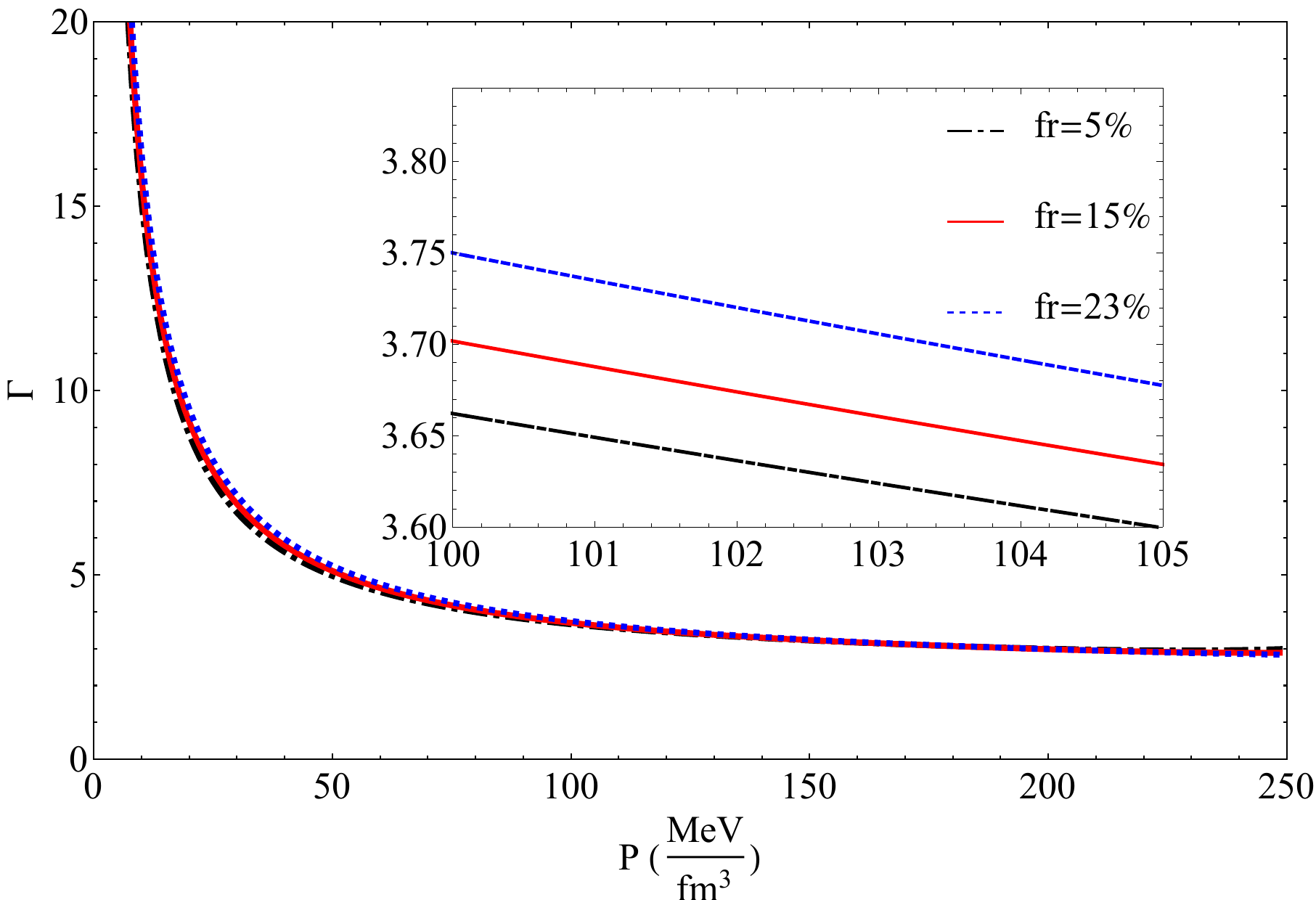}
		{$m_{\text{DM}}$=160 MeV}
	\end{subfigure}
	\caption{Adiabatic index for different values of $m_{\text{DM}}$ and $fr$.}
	\label{Adias}
\end{figure}

The stability of compact stars is investigated by calculating the adiabatic index, $\Gamma=\frac{dP}{d\epsilon}\frac{P+\epsilon}{P}$, which its value must exceed $4/3$ for a relativistic gas in hydrostatic equilibrium. 
If $\Gamma$ drops below this value, the star becomes unstable and collapses under its own gravity \cite{Chandrasekhar1964, Kuntsem1988}. Fig. \ref{Adias} shows that all computed EOSs meet this condition, ensuring the stars remain stable.

\section{structural properties of SQS}\label{SPOfSQS}

In this section, we investigate the structure characteristics of SQS based on the EOSs derived in the previous sections, emphasizing gravitational mass, associated radius, and $\Lambda$. Utilizing the  constraint $\Lambda_{1.4M\odot}<580$ from GW170817, we examine the maximum $M_{\text{TOV}}$ achievable by the EOSs for varying dark matter masses. Initially, we determine how the mass relates to the central energy density and the star's radius, followed by a discussion of its correlation with $\Lambda$.

\subsection{Mass and radius} 

\begin{figure}[h!]
	\centering
	\begin{subfigure}{0.45\textwidth}
		\centering
		\includegraphics[width=\textwidth]{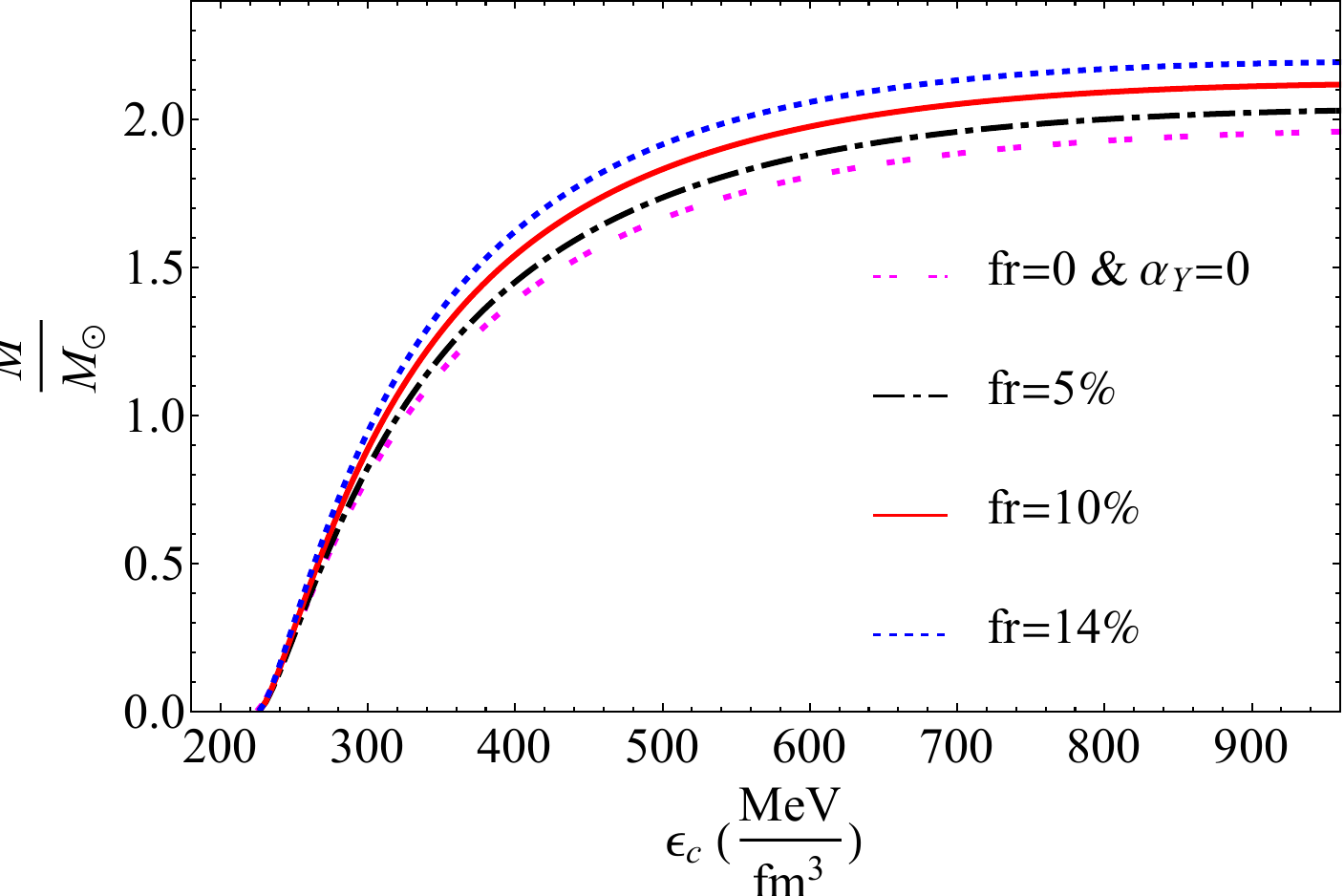}
		{$m_{\text{DM}}$=50 MeV}
	\end{subfigure}
	\hfill
	\begin{subfigure}{0.45\textwidth}
		\centering
		\includegraphics[width=\textwidth]{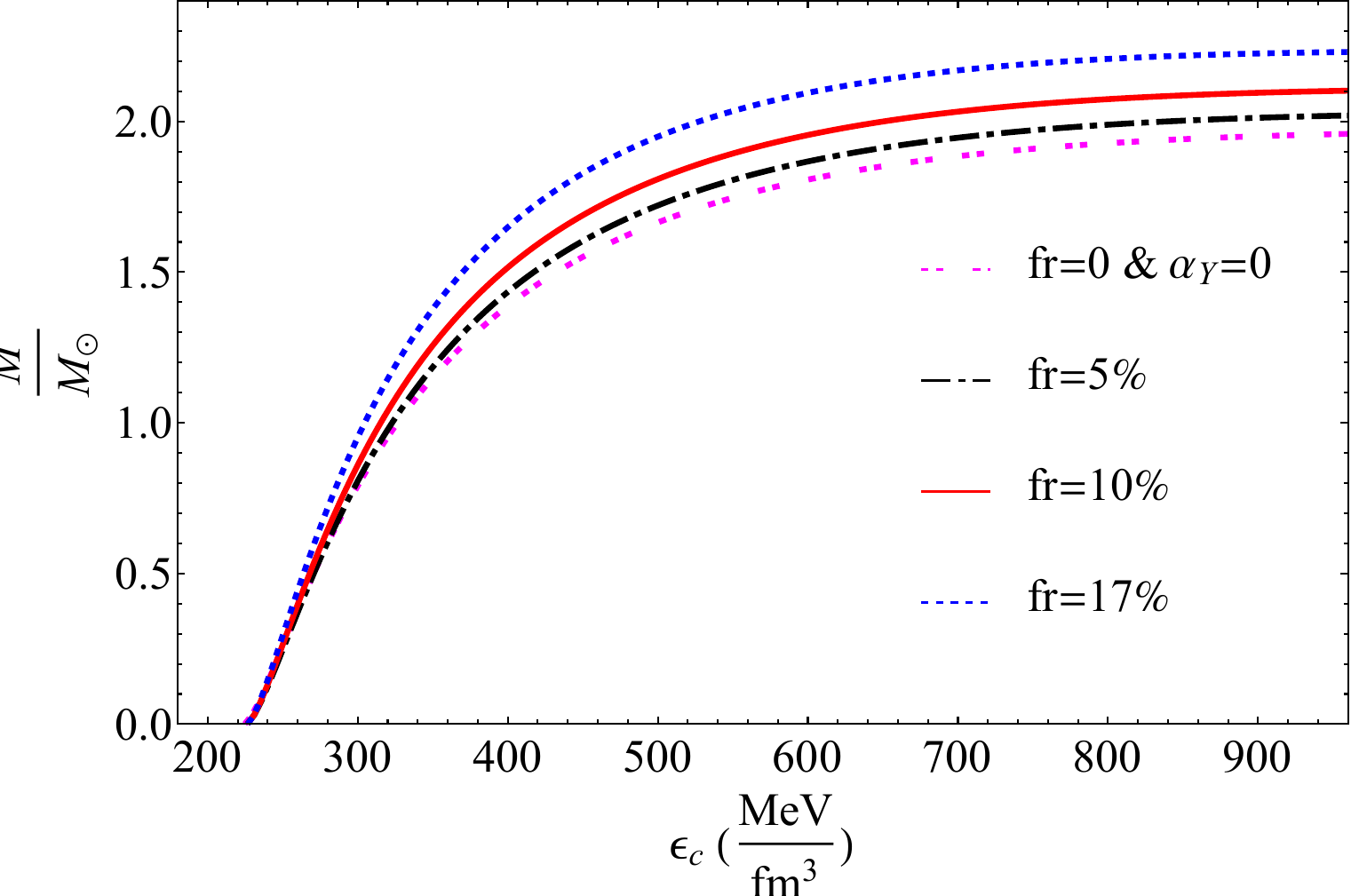}
		{$m_{\text{DM}}$=100 MeV}
	\end{subfigure}
	\hfill
	\begin{subfigure}{0.45\textwidth}
		\centering
		\includegraphics[width=\textwidth]{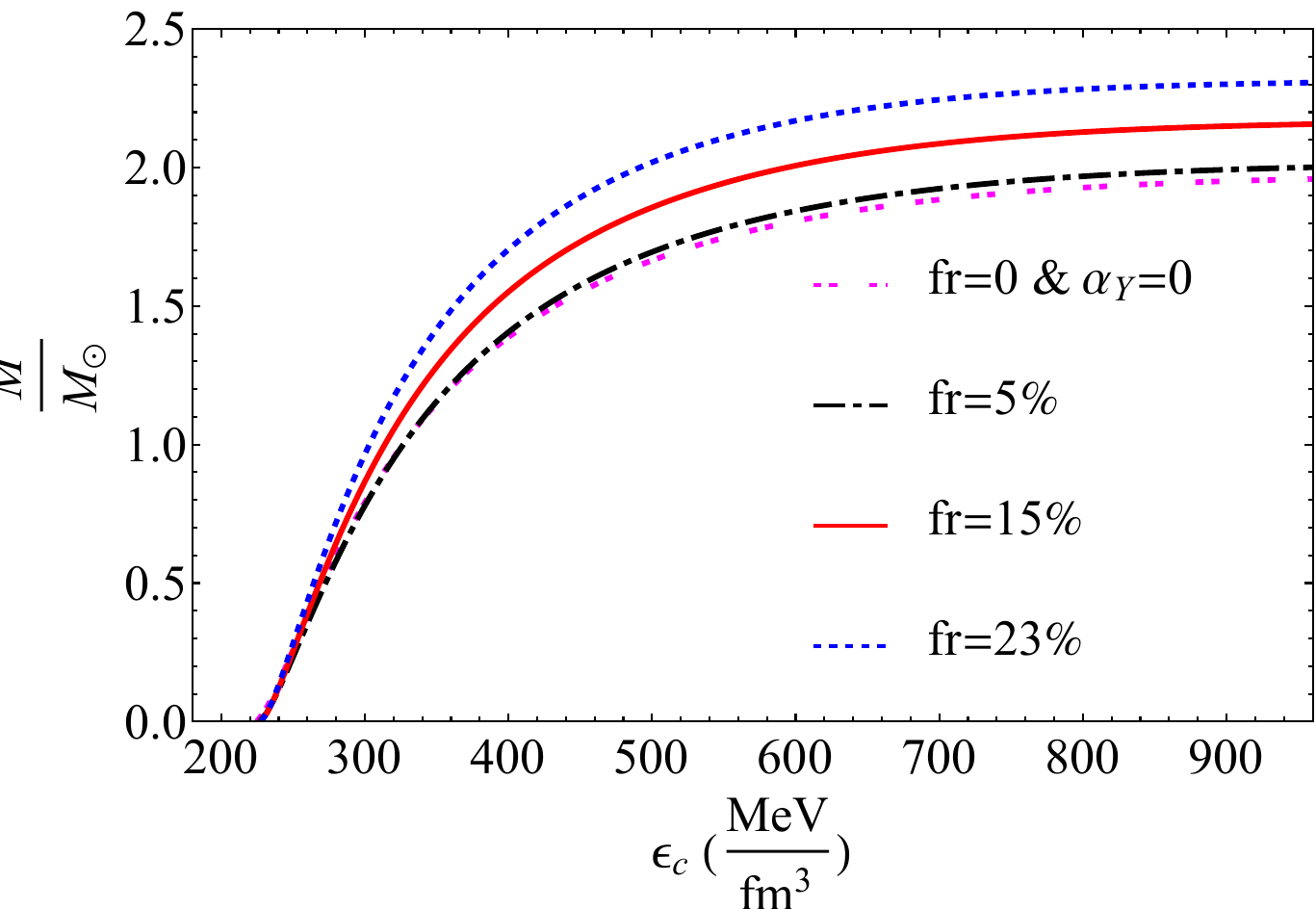}
		{$m_{\text{DM}}$=160 MeV}
	\end{subfigure}
	\caption{Mass as a function of central energy density for different values of $m_{\text{DM}}$ and $fr$.}
	\label{mes}
\end{figure}

\begin{figure}[h!]
	\centering
	\begin{subfigure}{0.45\textwidth}
		\centering
		\includegraphics[width=\textwidth]{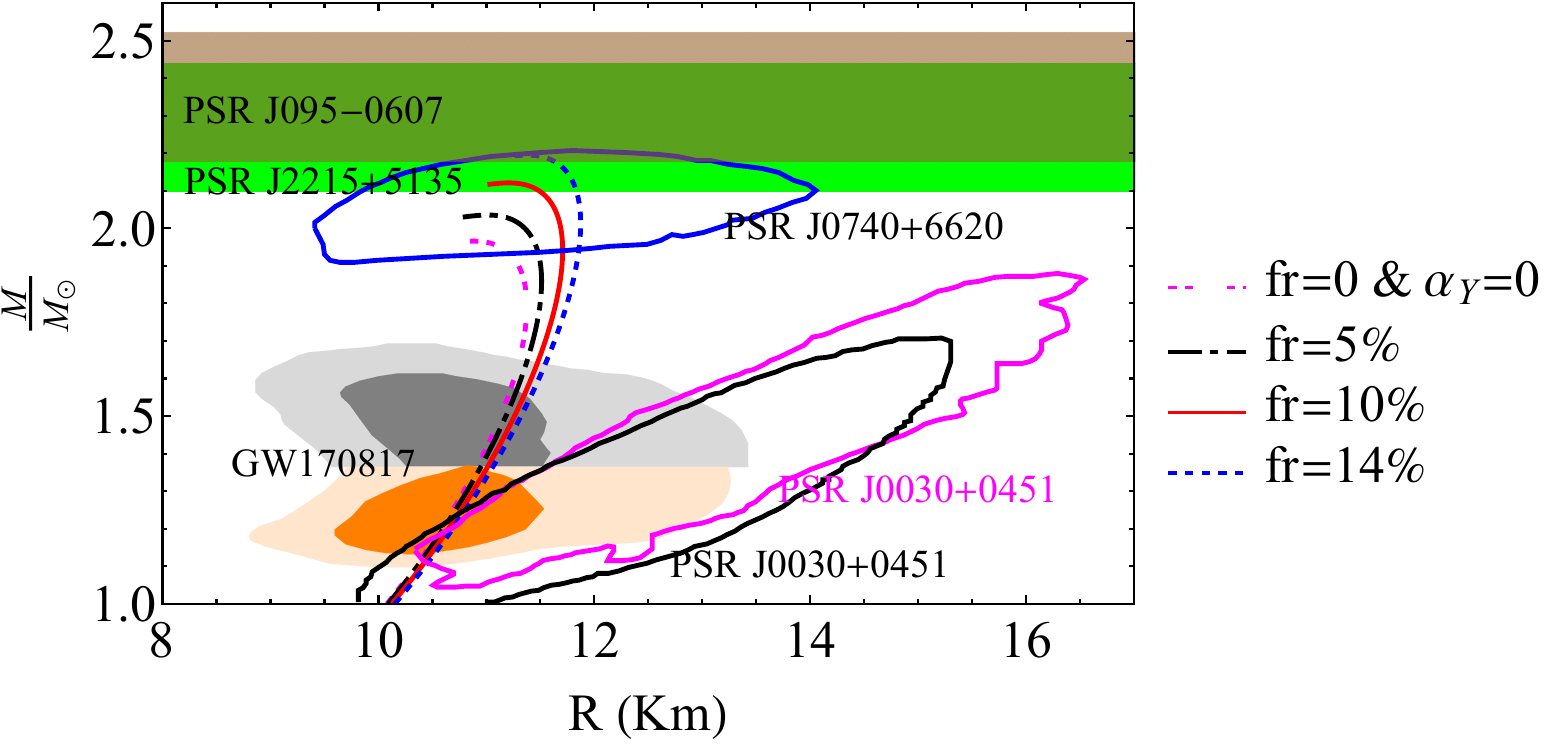}
		{$m_{\text{DM}}$=50 MeV}
	\end{subfigure}
	\hfill
	\begin{subfigure}{0.45\textwidth}
		\centering
		\includegraphics[width=\textwidth]{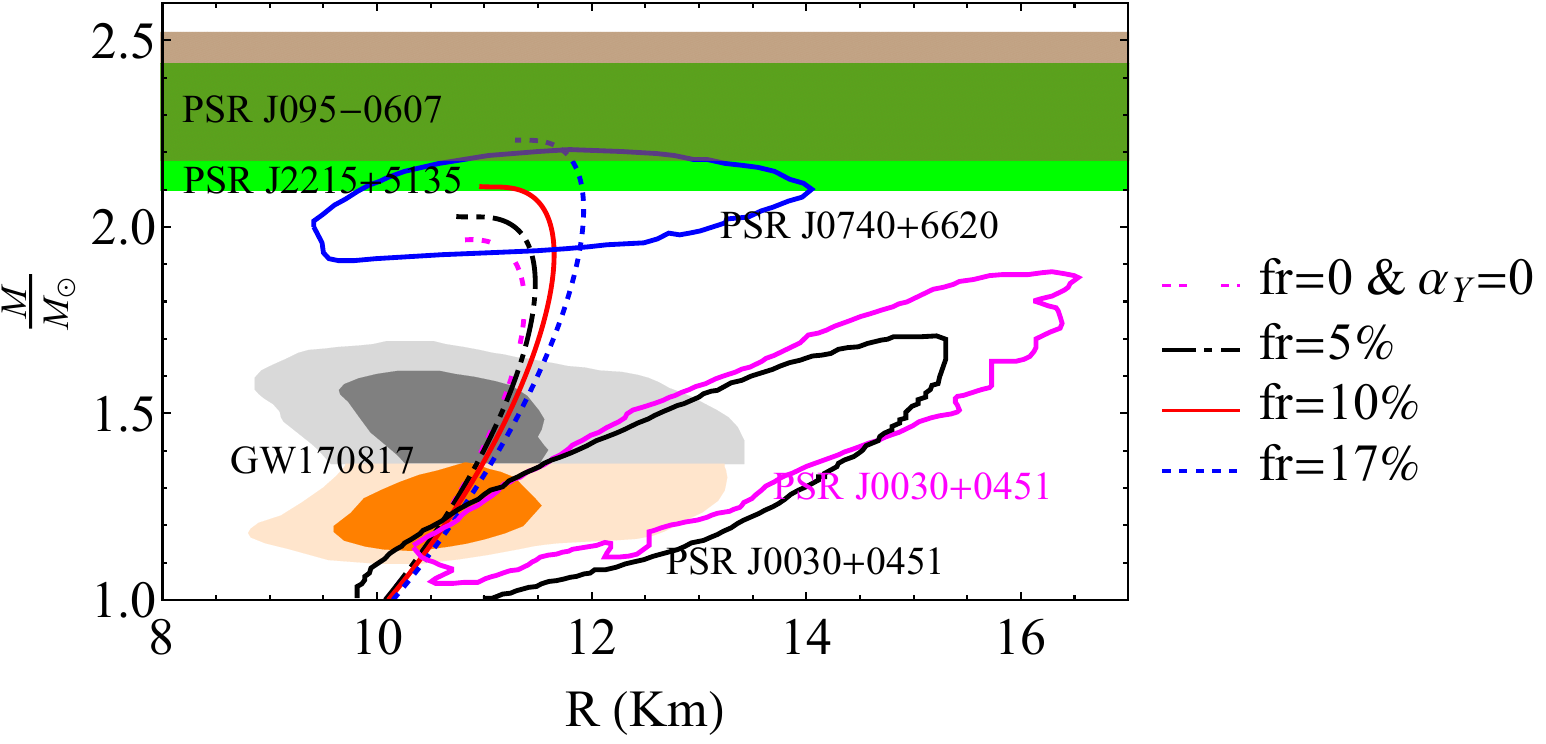}
		{$m_{\text{DM}}$=100 MeV}
	\end{subfigure}
	\hfill
	\begin{subfigure}{0.45\textwidth}
		\centering
		\includegraphics[width=\textwidth]{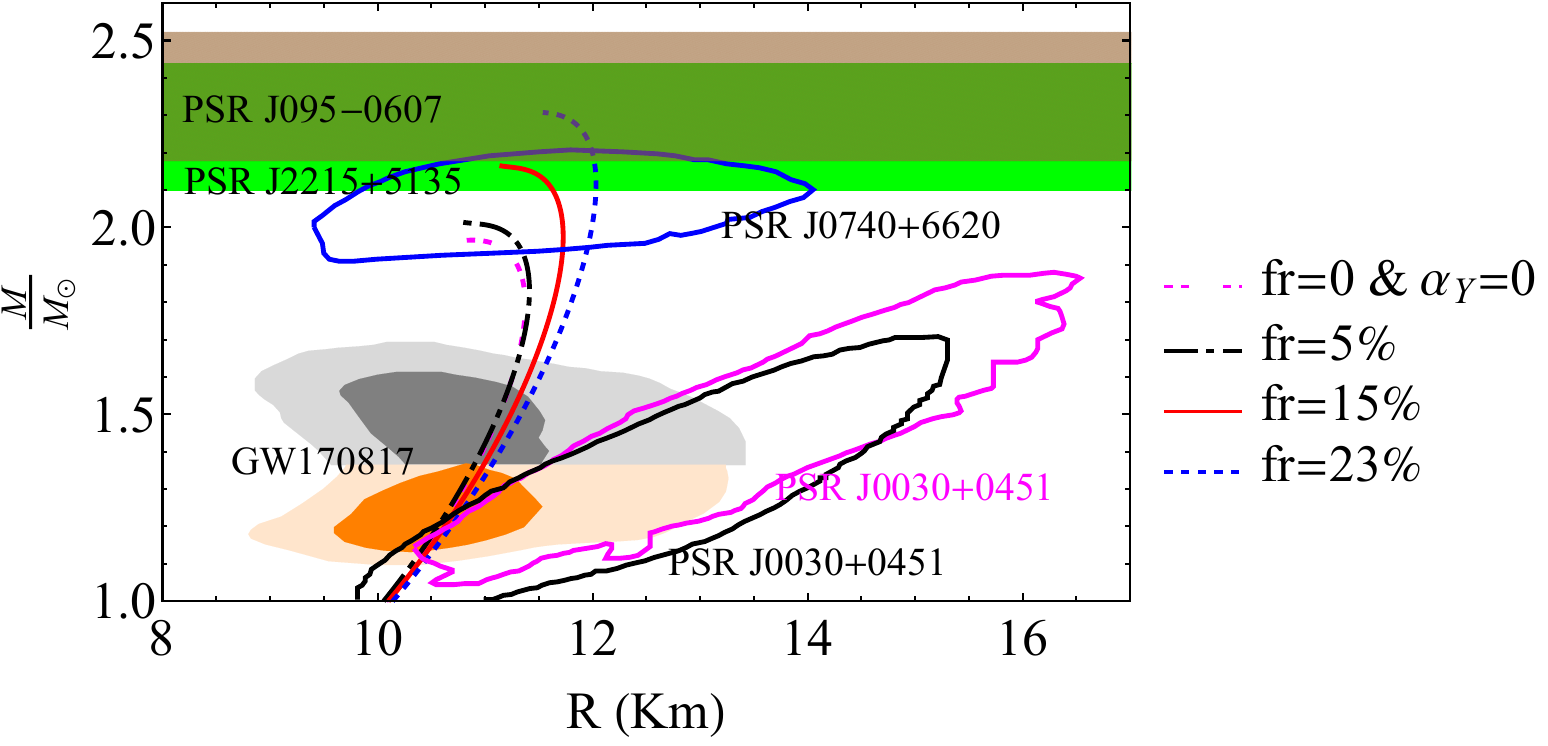}
		{$m_{\text{DM}}$=160 MeV}
	\end{subfigure}
	\caption{M-R diagrams for different values of $m_{\text{DM}}$ and $fr$.}
	\label{mrs}
	\end{figure}

The Tolman-Oppenheimer-Volkoff (TOV) equations describe the structure of a spherically symmetric, static star in hydrostatic equilibrium. These equations couple the pressure, mass, and density of the star to the spacetime geometry under the influence of general relativity. The equations are \cite{Tolman1939,Oppenheimer1939}:
\begin{equation}
M(r)=\int_{0}^{R}4\pi r^{2}\epsilon(r)dr,
\end{equation}
and
\begin{eqnarray}
\frac{dP(r)}{dr} &=&\dfrac{\left[ P(r)+\epsilon (r)\right] \left[ M(r)+4\pi
	r^{3}P(r)\right] }{r\left( 2M(r)-r\right) },\label{TOV}
\end{eqnarray} 
where $M(r)$ represents the enclosed mass within radius $r$, $\epsilon(r)$ denotes the energy density, and $P(r)$ is the pressure. We start by considering different central pressures based on the perturbative EOSs. The mass at the center of the star is initially zero, and we solve the equations until the pressure becomes zero at the surface. This gives us the star's mass and radius, along with the relationship between central pressure or energy density and gravitational mass. We repeat this process for various central pressures allowed by the EOSs. The results depicted in Fig. \ref{mes} present the gravitational mass as a function of the central energy density ($\epsilon_c$). In Fig. \ref{mes}, we have considered the case of a quark star without the presence of dark matter to gain a better understanding of the presence of dark matter and its impact on the star's mass. To ensure the stability of the star, the condition $\frac{\partial M}{\partial \epsilon_c} > 0 $ must be satisfied \cite{Tangphati2021a,Tangphati2021b}.
As shown in Fig. \ref{mes}, this condition is fulfilled for all the diagrams. The peak of the mass curve represents the maximum gravitational mass of the SQS where this condition holds true. The results clearly show that considering dark matter increases the maximum mass of the star. Moreover, an interesting result is that at the same $fr$ (for example at $fr=5\%$), increasing the $m_{\text{DM}}$ decreases the maximum mass of SQS. However, increasing the $fr$ at any given $m_{\text{DM}}$ leads to an increase in the star's maximum mass. Of course, such a result was predictable given the behavior of the speed of sound previously analyzed. Examining such a result is more evident in the mass-radius (M-R) plot. Now we proceed to obtain the M-R plots. Fig. \ref{mrs} illustrates the M-R  relationship for SQSs derived using our EOSs.
The diagrams includes key observational constraints: i) The brown bar represents the mass measurement of PSR J0952-0607, the heaviest known pulsar with a mass of $2.35 \pm 0.17 M_\odot$ \cite{J0952}.  Our EOS curves, corresponding to the maximum values of $ fr $ for each $ m_{\text{DM}} $, intersect this region, demonstrating their capability to describe ultra-massive compact stars. ii) The green bar marks the mass constraint from PSR J2215+5135, a high-mass pulsar with an estimated mass around $2 M_\odot$ \cite{Linares2018}. As we can see, the EOS for SQM without the presence of dark matter does not cover any of the massive masses such as PSR J0952-0607 and PSR J2215+5135. The alignment of our results (EOSs in presence of dark matter) with the range of J2215+5135 underscores the validity of our EOSs for describing stars in this mass range. iii) The gray and orange areas correspond to the constraints on the mass and radius from the gravitational wave event GW170817.  Our EOS curves lie within both regions, demonstrating consistency with the constraints imposed by this event. iv) The solid black and pink curves correspond to the radius constraints for PSR J0030+0451 as measured by NICER \cite{Miller2019,Riley2019}. The black curve shows the lower-radius boundary, while the pink curve represents the upper-radius boundary. Our M-R relation intersects this region aligning with the NICER results. v) The blue curve represents the radius constraints for PSR J0740+6620 derived from NICER and XMM-Newton observations for a high-mass star ($2.1 M_\odot$) \cite{Cromartie2019}. Our EOSs fit within this range, supporting their robustness for describing massive SQSs. The M-R curves computed using our EOSs capture the full range of observational constraints, from the masses of ultra-massive pulsars to radius measurements limits. The curves exhibit a balance between stiffness (supporting high masses) and compactness (satisfying radius constraints), making it a viable model for SQSs. Fig. \ref{mrs} illustrates that an increase in $fr$, leads to a noticeable increase in the maximum mass (TOV mass). This enhancement directly reflects the role of dark matter in stiffening the EOS, allowing the star to support higher masses. In the following, we present the behavior of $\Lambda$ as a function of the mass of SQS for different values of $m_{\text{DM}}$ and various $fr$. 

\subsection{Tidal deformability}

The studying SQSs offers a unique window into the physics of dense matter and potential deviations from standard nuclear EOSs. Among the astrophysical observables that constrain the properties of compact stars, the $\Lambda$ parameter, $\Lambda$, has gained significant importance following the detection of GWs from binary neutron star mergers. This dimensionless parameter encodes information about the star's internal structure and its response to the tidal forces in a binary system, serving as a critical tool for probing exotic states of matter, such as those posited for SQS. We utilized the constraint derived from the dimensionless $\Lambda$ to determine the maximum $fr$ of dark matter within an SQS. To compute $\Lambda$, the metric function $H(r)$ must be determined using the following equation \cite{Hinderer,ChengMingLi2020,Sedaghatepjc2024},
\begin{eqnarray}
	\frac{d\beta}{dr} &=& 2(1-2\frac{m_r}{r})^{-1}H\{-2\pi[5\epsilon+9p+f(\epsilon+p)]\nonumber\\
	&+&\frac{3}{r^2}+2(1-2\frac{m_r}{r})^{-1}(\frac{m_r}{r^2}+4\pi rp)^2\}\nonumber\\
	&+&\frac{2\beta}{r}(1-2\frac{m_r}{r})^{-1}\{\frac{m_r}{r}+2\pi r^2(\epsilon-p)-1\}\,\label{HbetaEq},
\end{eqnarray}
where, $\beta = \frac{dH}{dr}$ and $f$ denotes $\frac{d\epsilon}{dp}$.  The parameter $\Lambda $ is then obtained by the formula:
\begin{equation}
	\Lambda =\frac{2}{3}k_{2}R^{5}.
\end{equation}
Here, $k_{2}$ represents the dimensionless tidal Love number for $l = 2$. The expression for $k_{2}$ is given by:
\begin{align}
	k_{2}& =\frac{16\sigma ^{5}}{5}(1-2\sigma )^{2}\left[ 1+\sigma (y-1)-\frac{y%
	}{2}\right]  \notag \\
	& \times \left\{ 12\sigma \left[ 1-\frac{y}{2}+\frac{\sigma (5y-8)}{2}\right]
	\right.  \notag \\
	& +4\sigma ^{3}\left[ 13-11y+\sigma \left( 3y-2\right) +2\sigma ^{2}\left(
	1+y\right) \right]  \notag \\
	& +\left. 6(1-2\sigma )^{2}\left[ 1+\sigma (y-1)-\frac{y}{2}\right] \ln
	\left( 1-2\sigma \right) \right\} ^{-1/2},  \label{tln}
\end{align}
where, $\sigma$ is defined as $M/R$, and $y$ is expressed as:
\begin{equation}
 y=\frac{R \beta(R)}{H(R)} - \frac{4\pi R^3 \epsilon_0}{M}.
\end{equation}  
In above relation, $\epsilon_0$ represents the energy density at the surface of the SQS \cite{ChengMingLi2020,Sedaghatepjc2024}.
By simultaneously solving Eqs. (\ref{TOV}) and (\ref{HbetaEq}), along with $\frac{dM}{dr} = 4\pi r^2 \epsilon$, we obtain the relationship that expresses $\Lambda$ as a function of mass. We now analyze our results in light of the constraint on $\Lambda$ obtained from the binary system GW170817.
\begin{figure}[h!]
	\centering
	\begin{subfigure}{0.45\textwidth}
		\centering
		\includegraphics[width=\textwidth]{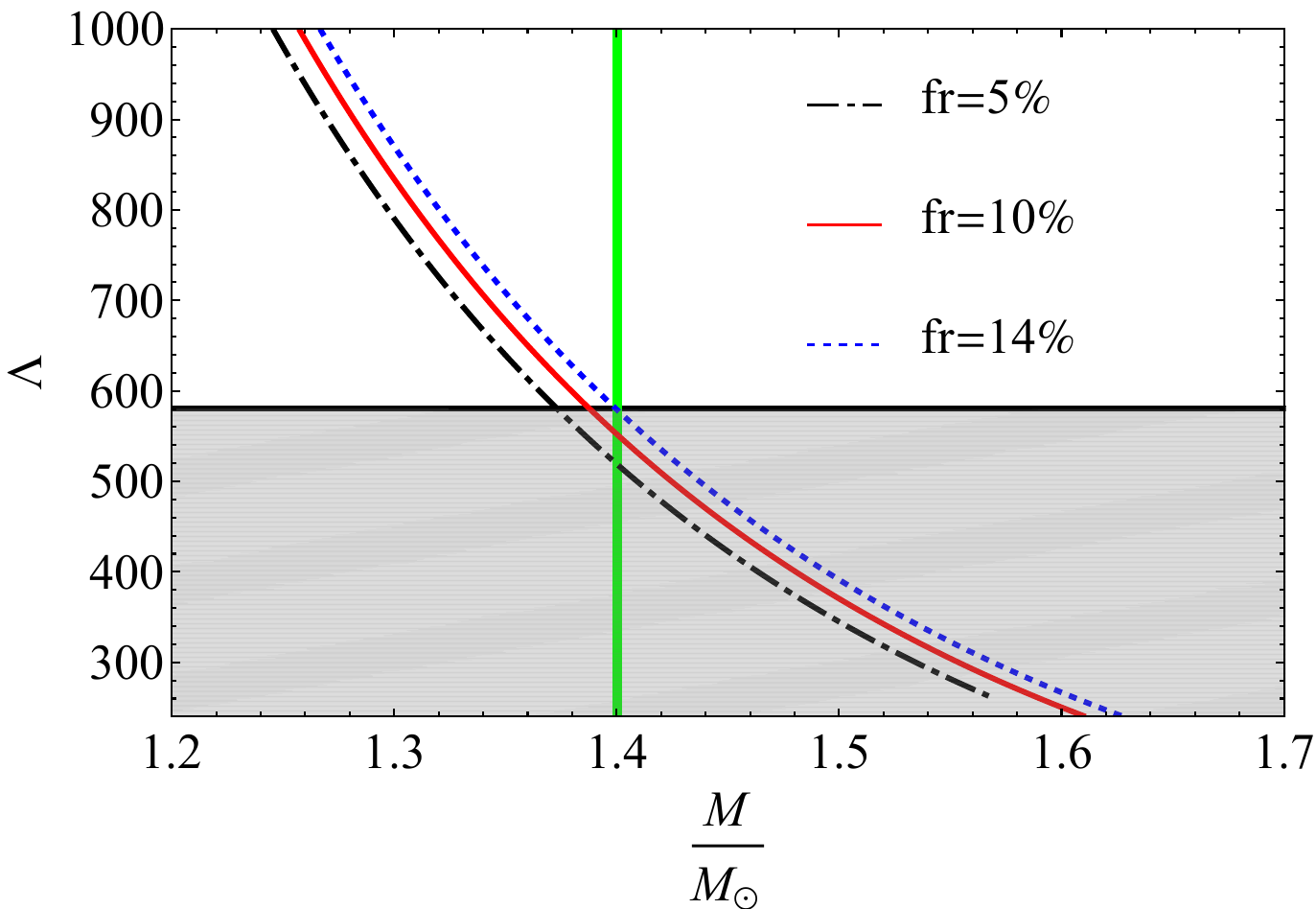}
		{$m_{\text{DM}}$=50 MeV}
	\end{subfigure}
	\hfill
	\begin{subfigure}{0.45\textwidth}
		\centering
		\includegraphics[width=\textwidth]{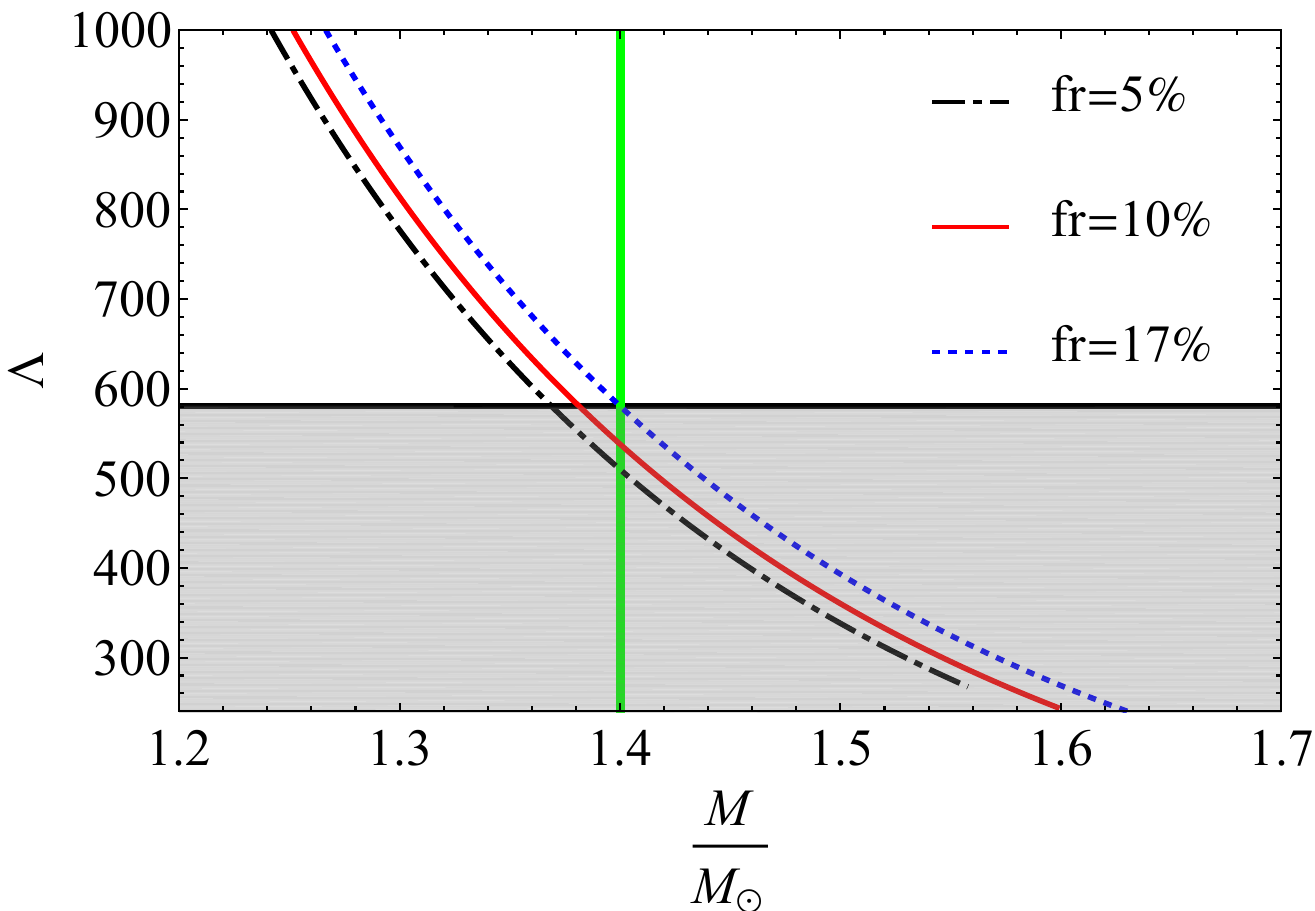}
		{$m_{\text{DM}}$=100 MeV}
	\end{subfigure}
	\hfill
	\begin{subfigure}{0.45\textwidth}
		\centering
		\includegraphics[width=\textwidth]{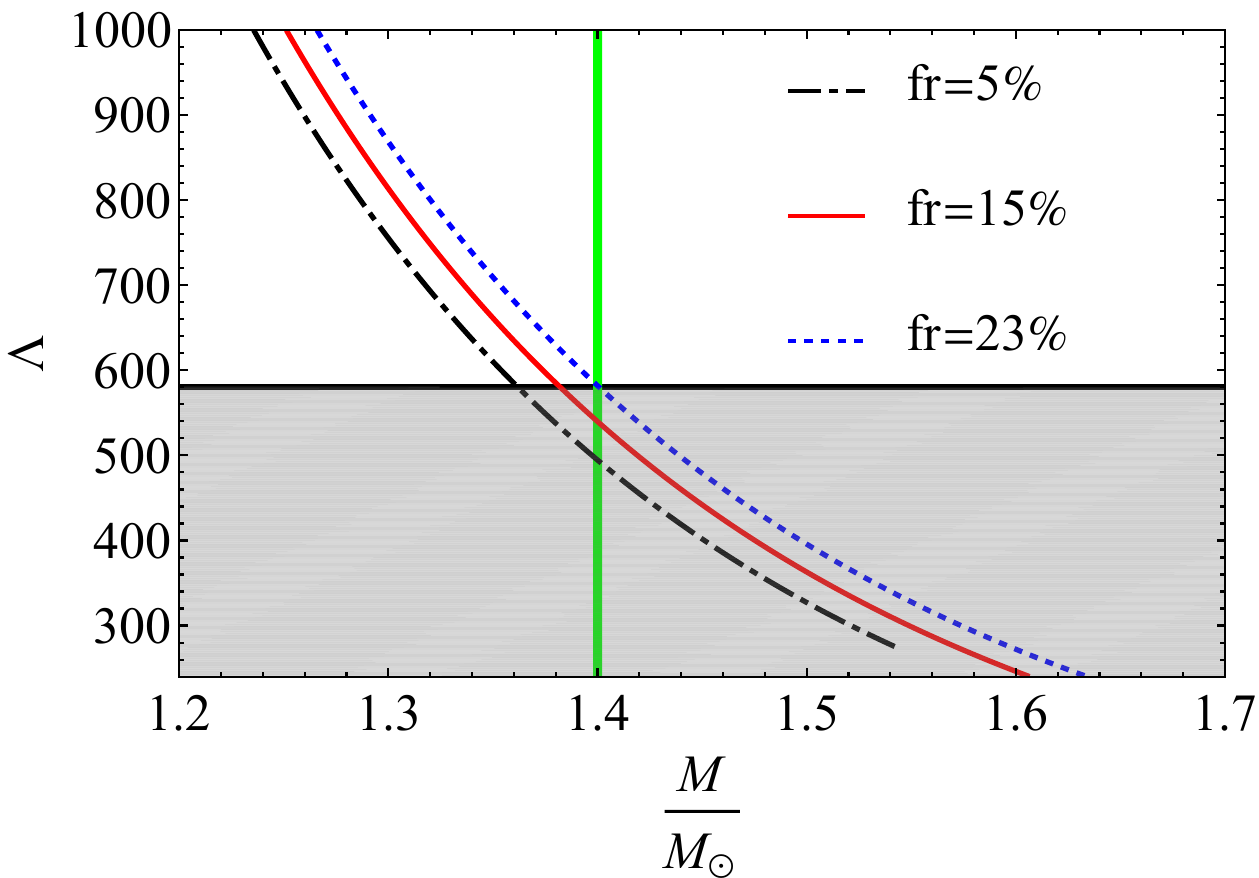}
		{$m_{\text{DM}}$=160 MeV}
	\end{subfigure}
	\caption{Dimensionless tidal deformability as a function of mass for different values of $m_{\text{DM}}$ and $fr$.}
	\label{tidals}
\end{figure}
Fig. \ref{tidals} show the behavior of $\Lambda$ versus mass of SQS for different values of $m_{\text{DM}}$ and various $fr$. The maximum $fr$ is chosen to ensure the constraint $\Lambda_{1.4M\odot}<580$  is respected. This is represented by the black horizontal line in each figure. Increasing the $fr$ stiffens the EOS, resulting in higher $\Lambda$. On the other hand, increasing the mass of dark matter softens the EOS, leading to a reduction in $\Lambda$ for a fixed $fr$. More clearly, an increase in $m_{\text{DM}}$ reduces the contribution of BEC of dark matter, resulting in a softer EOS, which increases the possibility of star deformability. As a result, the maximum allowable $fr$ rises with increasing dark matter mass. The reasons for this behavior have been discussed in detail in section \ref{EOSSA}. Table \ref{results} presents the structural properties of SQS for various values of $m_{\text{DM}}$ and  $fr$. $fr_{max}$ represents the maximum $fr$ in SQS that satisfies the constraint $\Lambda_{1.4M\odot}<580$. The last two columns of table \ref{results} show the contribution of dark matter to the TOV mass of SQS. As the results indicate, this contribution increases with the rise in dark matter mass. For dark matter with a mass $50$ MeV, the contribution is below one percent, and for dark matter with a mass of $100$ MeV., it exceeds one percent only for a $fr$ of $17\%$. The maximum contribution, about 3.46\%, occurs for dark matter with a mass of $160$ MeV and a $fr$ of 23\%.
\begin{table}[h!]
	\caption{		
		Structural properties of SQS for different values of $m_{\text{DM}}$ and $fr$, under the $\Lambda$ constraint from GW170817.}
	\centering
	\small
	\begin{subtable}{1.0\linewidth}
		\centering
		\scalebox{1.0}{
			\begin{tabular}{|c|c|c|c|c|c|c|}
				\hline
				\multicolumn{6}{|c|}{\textbf{$m_{\text{DM}}$=50 MeV}} \\
				\hline
				$fr$ & $\Lambda_{1.4\textup{M}_\odot}$& $R(km)$ & $M_{\text{TOV}}(\textup{M}_\odot)$ & $M_{\text{DM}} (\textup{M}_\odot)$& $\frac{M_{\text{DM}}}{M_{\text{TOV}}}$(\%)  \\\hline
				$5\%$ & 517.69 & 10.89 & 2.03 & 0.005 & 0.25\\ \hline
				$10\%$ & 552.17 & 11.22 & 2.12 & 0.008 & 0.38 \\ \hline
				$14\%$ & 578.67 & 11.31 & 2.19 & 0.01 & 0.46 \\ \hline
			\end{tabular}}
			
		\end{subtable}
		\\
	\begin{subtable}{1.0\linewidth}
		\centering
		\scalebox{1.0}{
			\begin{tabular}{|c|c|c|c|c|c|c|}
				\hline
				\multicolumn{6}{|c|}{\textbf{$m_{\text{DM}}$=100 MeV}} \\
				\hline
				$fr$ & $\Lambda_{1.4\textup{M}_\odot}$& $R(km)$ & $M_{\text{TOV}}(\textup{M}_\odot)$ & $M_{\text{DM}} (\textup{M}_\odot)$ & $\frac{M_{\text{DM}}}{M_{\text{TOV}}}(\%)$ \\\hline
				$5\%$ & 509.94 & 10.88 & 2.02 & 0.01 & 0.5\\ \hline
				$10\%$ & 537.53 & 10.95 & 2.11 & 0.02 & 0.95\\ \hline
				$17\%$ & 579.20 & 11.28 & 2.23 & 0.03 & 1.34\\ \hline
			\end{tabular}}
			
		\end{subtable}
			\\
	\begin{subtable}{1.0\linewidth}
		\centering
		\scalebox{1.0}{
			\begin{tabular}{|c|c|c|c|c|c|c|}
				\hline
				\multicolumn{6}{|c|}{\textbf{$m_{\text{DM}}$=160 MeV}} \\
				\hline
				$fr$ & $\Lambda_{1.4\textup{M}_\odot}$& $R(km)$ & $M_{\text{TOV}}(\textup{M}_\odot)$ & $M_{\text{DM}}(\textup{M}_\odot)$ & $\frac{M_{\text{DM}}}{M_{\text{TOV}}}$(\%) \\\hline
				$5\%$ & 492.84 & 10.76 & 2.01 & 0.01 &  0.5\\ \hline
				$15\%$ & 539.88 & 11.13 & 2.16 & 0.05 & 2.31\\ \hline
				$23\%$ & 578.99 & 11.53 & 2.31 & 0.08 & 3.46\\ \hline
			\end{tabular}}
			
		\end{subtable}
				\label{results}
			\end{table}	

\section{conclusion}
This study investigated the influence of scalar dark matter on SQS using a one-fluid model, accounting for Yukawa interactions between dark matter and quark matter. {The SQM consists of up and down quarks with zero mass and a strange quark with a running mass. To describe the running behavior of the strange quark mass and QCD coupling as functions of energy, we utilized the latest results from the particle data group \cite{Workman}.} The model incorporates contributions from QCD interactions, Yukawa interaction between dark matter and quark, and Bose-Einstein condensation (BEC) of dark matter, presenting a comprehensive approach to understanding the structural properties of SQS in the presence of dark matter.
We first derived the thermodynamic potential by combining free quark, QCD, Yukawa, and BEC contributions. This formulation allowed us to capture the effects of both perturbative QCD and dark matter interactions on the EOS. {To derive the thermodynamic potential from QCD interactions, we used a two-loop Feynman diagram, as performed in \cite{Fraga2006} and \cite{Kurkela2010}. For the Yukawa interaction between dark matter and quark matter, we calculated the amplitude of a two-loop Feynman diagram that includes the dark matter mass ($m_{\text{DM}}$) and the Yukawa coupling constant, $\alpha_Y$. 
The values of $m_{\text{DM}}$ were chosen based on constraints on the dark matter self-interaction cross-section from Refs. \cite{Panotopoulos2017b,Lopes2018}, using three specific values: $50$ MeV, $100$ MeV, and $160$ MeV. To determine the allowable range of $\alpha_Y$, we applied the stability condition for SQM, ensuring the energy per baryon stayed below the stability threshold. For each $m_{\text{DM}}$ value, we identified the maximum allowable $\alpha_Y$. Our calculations revealed that increasing \( m_{\text{DM}} \) also raises the maximum permissible value of \( \alpha_Y \). Finally, we selected a fixed value of $\alpha_Y$ (0.1) that satisfied the stability condition across all $m_{\text{DM}}$ values. Using this value of $\alpha_Y$, we derived the EOS, the speed of sound, and the adiabatic index for various dark matter fractions ($fr$) contributing to the total pressure of SQS. These fractions were constrained by the tidal deformability limit from GW170817.} 
We verified that the EOSs satisfy fundamental physical constraints such as causality and dynamical stability.  {The results showed that increasing the $fr$ stiffens the EOS, leading to higher sound speeds.  Conversely, increasing the $m_{\text{DM}}$ softens the EOS at a fixed $fr$, resulting in slightly lower sound speeds.}
In the following, we investigated the structure properties of SQS, including their mass-radius relationship, tidal deformability, and maximum gravitational mass. {We carried out calculations for the structure of the SQS both with and without dark matter. In the absence of dark matter, we found that EOS from perturbative QCD is unable to support massive objects as large as $2M_{\odot}$.} The incorporation of dark matter significantly altered the star's structure properties. {We investigated these properties for different values of $m_{\text{DM}}$ and  $fr$. The calculations were constrained by the tidal deformability limit from GW170817.} {We showed that at a constant $m_{\text{DM}}$, increasing the $fr$ enhances the BEC contribution, producing a more massive SQS. Conversely, increasing $m_{\text{DM}}$ at a fixed $fr$ reduces the SQS mass.} The reason for this behavior were discussed in detail in section \ref{EOSSA}. We showed that our EOSs successfully describe ultra-massive pulsars such as PSR J0952-0607 and PSR J2215+5135, which cannot be explained by QCD-based EOSs alone. \textcolor{black}{We also evaluated the impact of dark matter on the TOV mass of SQS. For dark matter with a mass of $50$ MeV, the contribution is below one percent. At a mass of $100$ MeV, it can exceed 1.34\%. The maximum contribution, around 3.46\%, occurs when the dark matter mass reaches $160$ MeV.}
By adding dark matter into the EOS of SQM, our study connects theoretical models to astrophysical observations. Future studies could extend this framework to explore additional interactions, alternative dark matter candidates, or upcoming observations from GW detectors.
\section*{Acknowledgements}
We wish to thank Shiraz University Research Council. This work is based upon research funded by Iran National Science Foundation (INSF) under project No. 4022870.	

\end{document}